\documentclass[prc,aps,superscriptaddress,floatfix,amsmath,amssymb,floatfix,nofootinbib]{revtex4}
\usepackage[plainpages=false, colorlinks=true, anchorcolor=lblui, linkcolor=blue, citecolor=black, urlcolor=blue, bookmarks=false]{hyperref}
\usepackage[usenames]{color}
\usepackage{soul}
\usepackage{mathtools}
\usepackage{amssymb}
\usepackage{feynmp}
\usepackage{amsmath}
\usepackage{amssymb}
\usepackage{graphicx}
\usepackage{epsfig}
\usepackage{mdframed}
\usepackage{mathtools}
\ifpdf                                     % For pdfLaTeX
  \usepackage{hyperref,graphicx}           %   usepackage without driver option
\else                                      % For (p)LaTeX + dvipdfmx
  \usepackage[dvipdfmx]{hyperref,graphicx} %   usepackage with driver option
  \usepackage{mediabb}                     %   [For PDF graphics] Automatic PDF-bounding box
\fi

\newcommand{\script}[1]{\begin{cal} #1 \end{cal} }

\newcommand{\Pol}{\boldsymbol{\Pi}}
\newcommand{\be}{\begin{equation}}
\newcommand{\ee}{\end{equation}}
\newcommand{\bea}{\begin{eqnarray}}
\newcommand{\eea}{\end{eqnarray}}
\newcommand{\ba}{\begin{array}}
\newcommand{\ea}{\end{array}}

\newcommand{\code}[1]{\verb~#1~}

\newcommand{\fsl}[1]{\ensuremath{\mathrlap{\!\not{\phantom{#1}}}#1}}
\newcommand{\codewebsite}{\texttt{\url{https://bitbucket.org/lroberts/nuopac}}}

\long\def\symbolfootnote[#1]#2{\begingroup%
\def\thefootnote{\fnsymbol{footnote}}\footnote[#1]{#2}\endgroup}

\newcommand{\im}[1]{\textrm{Im}#1}
\newcommand{\re}[1]{\textrm{Re}#1}
\newcommand{\muhat}[1]{\hat{\mu}}

\newcommand{\newacronym}[3]{%
 \newcommand{#1}[1]{#3##1 (#2##1)%
 \renewcommand{#1}[1]{#2####1}}%
}

\newcommand{\pp}[0]{I}

\newacronym{\EOS}{EoS}{equation of state}
\newacronym{\RPA}{RPA}{random phase approximation}

\def\apjl{ApJL }

\def\apj{ApJ }
\def\apjs{ApJ  Supplement}

\def\apjs{ApJS }

\def\aap{A\&A }

\usepackage{hyperref}

\begin{document} 
% INT-PUB-16-048
{\begin{flushleft} INT-PUB-16-048
\vskip -0.5in
{\normalsize }
\end{flushleft}
\title{Charged current neutrino interactions in hot and dense matter}
\author{Luke F. Roberts}
\affiliation{National Superconducting Cyclotron Laboratory and Department of
Physics and Astronomy, Michigan State
University, East Lansing, Michigan 48824, USA}
\email{robertsl@nscl.msu.edu}
\author{Sanjay Reddy}
\affiliation{Institute for Nuclear Theory, University of Washington, 
Seattle, Washington 98195, USA}
\email{sareddy@uw.edu}
%======================================================================
%%%%%%%%%%%%%%%%%%%%%%%%%%%%%%%%%%%%%%%%%
\begin{abstract}
We derive the charged current absorption rate of electron and anti-electron
neutrinos in dense matter using a fully relativistic approach valid at
arbitrary matter degeneracy.  We include mean field energy shifts due to
nuclear interactions   and the corrections due to weak magnetism. The rates are
derived both from the familiar Fermi's Golden Rule, and from the techniques of
finite temperature field theory, and shown to be equivalent. In various limits,
these results can also be used to calculate neutral current opacities.  We find
that some pieces of the response have been left out in previous derivations and
their contribution at high density can be significant.  Useful formulae and
detailed derivations are presented and we provide a new open source
implementation of these opacities for use in radiation hydrodynamic simulations
of core-collapse supernovae and neutron star mergers.

\end{abstract}
%%%%%%%%%%%%%%%%%%%%%%%%%%%%%%%%%%%%%%%%%

%\keywords{neutron stars, radiative transfer, neutrinos} 
\maketitle 

\section{Introduction} 

Neutrino opacities in dense matter are of paramount importance to the evolution of core-collapse
supernovae and the remnants of compact object mergers. They impact the
properties of the neutrino signal of these events \cite[e.g.][]{Pons:99,
Hudepohl:10}, are integral to the rate of energy transport
\cite[e.g.][]{Janka:16}, and can strongly alter the composition of matter
ejected in these events \cite[e.g.][]{Wanajo:14, Radice:16}. 
Recent work on modeling core-collapse supernovae has shown that
three-dimensional models are close to explosion, but the results are sensitive
to small changes in the neutrino opacities and other simulation inputs \cite{Melson:15,
Burrows:16, Horowitz:16, Roberts:16a}. Therefore, it is important to
provide accurate expressions for the neutrino opacities required for these
numerical models.  

In dense matter many-body effects can modify the neutrino mean free paths.  The
inclusion of the nucleon self energies and effective masses in the medium can
significantly alter electron neutrino and anti-neutrino emission. These effects
were first realized in \cite{Reddy:98} and have been the focus of recent work
because of its implications for neutrino spectra and nucleosynthesis
\cite{Martinez-Pinedo:12, Roberts:12a, Roberts:12b, Horowitz:12, Rrapaj:15}.  In
addition, earlier studies have shown that neutrino interactions in dense matter
are influenced by matter degeneracy, and strong and electromagnetic correlations
between nucleons and leptons in the dense medium \cite{Sawyer:75, Iwamoto:82,
Horowitz:91, Burrows:98, Reddy:99, Burrows:99}.  These effects can suppress the
neutrino opacity at and above nuclear saturation density \citep{Burrows:98,
Reddy:99, Horowitz:03} and significantly accelerate  protoneutron star cooling
at late times \citep{Hudepohl:10, Roberts:12}.  Multi-particle excitations
\citep{Lykasov:08, Roberts:12b} and coherent scattering from a mixed phase
\citep{Reddy:00, Sonoda:07, Horowitz:16a} may also strongly impact neutrino
interaction rates and neutrino emission in supernovae.  The strength of all of
these effects depend on the assumed form of the nuclear interaction which also
influences the nuclear equation of state (EoS). These effects on the EoS and
neutrino opacities are correlated \cite{Sawyer:75,Iwamoto:82, Horowitz:91,
Burrows:98,Reddy:99, Burrows:99} and they should be calculated from the same
nuclear interaction.

Charged current neutrino interaction rates, such as $\nu_e + n \rightarrow e^- +
p$ are particularly sensitive to changes in the dispersion relations of nucleons
because the potential energy difference between the proton and neutron can alter
the lepton kinematics, which are often frustrated by final state blocking due to
degeneracy \citep{Reddy:98, Roberts:12b, Martinez-Pinedo:12, Rrapaj:15}. Although there has
been significant focus on including these effects in neutrino interaction rates,
there is still no complete, relativistic formalism available. The work of
\cite{Burrows:99} completely neglects the impact of the potential energy
difference of neutrons and protons on the vector and axial response of the
nuclear medium.  While the work \cite{Reddy:99} includes some impacts of the
potential difference, it also neglects some aspects: 1) It does not include the
impact of the potential difference on the hadronic part of the weak interaction
matrix element. At high densities and/or large neutrino energies, these missing
terms may impact neutrino charged current mean free paths. 2) It assumes that
the nucleon weak charged vector current is conserved, but is not
the case because of differences between neutron and proton masses and their dispersion relations in dense matter \citep{Leinson:01}. This will result in a different structure for the
charged current mean field polarization tensor.  The correct inclusion of these
extra terms is likely to impact the response of the medium when correlations are
included through the RPA (see below). 3) They neglected weak magnetism
corrections, which can be important for predicting the difference between
electron neutrino and electron anti-neutrino spectra and nucleosynthesis in the
neutrino driven wind, as well as the deleptonization rates of protoneutron stars
\citep{Horowitz:02}.

As a base line for future studies that would include
correlations, we derive for the first time the charged current absorption rates for electron
neutrinos which include all of the following effects: 1) different mean-field potential
energy shifts for neutrons and protons in neutron-rich matter; 2) relativistic
contributions to the nucleon charged currents; 3) weak magnetism; and 4) effects
due to the violation of the isospin symmetry, and consequently the lack of
conservation of the nucleon charged current in asymmetric matter
\cite{Leinson:01}. We provide derivations of these results both from
the perspective of Fermi's Golden Rule, and in the language of finite temperature
quantum field theory. In the neutral current limit, these expressions reduce to
those given in \cite{Horowitz:03} (up to a sign in one part of the
tensor-tensor polarization function). A library for calculating neutrino
interaction rates based on this work is available at \codewebsite.

The paper is structured as follows:
In section \ref{sec:cc_opacity}, we derive the general form of the charged
current opacity from Fermi's Golden Rule. In section \ref{sec:polarization}, we
calculate the full charged current polarization tensor and show that its
imaginary part agrees with the Fermi's Golden Rule results. We then present
practical representations of the response in section
\ref{sec:practical}. We also discuss some
approximations to the charged current absorption rate in section
\ref{sec:approximations}. In section
\ref{sec:results}, we present limiting forms of the rates and assess the impact
of the new terms. Throughout, we set $\hbar = c = k_B = 1$ and use a metric with
signature $(+---)$.

\section{Charged Current Opacity} 
\label{sec:cc_opacity} 

The charged current interaction at low energies is described by the Fermi weak interaction Lagrangian 
\begin{equation}
{\cal L}=\frac{C G_F}{\sqrt{2}}~l_\mu~j_{\rm cc}^\mu\,,
\label{eq:Lweak}
\end{equation}
where $l_\mu=\bar{l}\gamma_\mu(1-\gamma_5)\nu$ is the lepton charged current,
$C = \cos \theta_c$ is the cosine of the Cabibbo angle, and 
\begin{equation}
j^\mu_{\rm cc}=\bar{\Psi}_p~\left( \gamma^\mu(g_V-g_A\gamma_5)+ F_2 \frac{i
\sigma^{\mu\alpha} q_\alpha}{2 M} \right)~\Psi_n
\label{eq:Jweak}
\end{equation}
 is the nucleon charged current which includes the vector, axial vector, and
 weak magnetism contributions, characterized by coupling strengths $g_V=1$,
 $g_A=1.26$, and $F_2=3.71$, respectively, and $M=(M_n+M_p)/2=938.9$ MeV and
 $M_p,M_n$ are the proton and neutron masses, respectively. Here, the currents
 are written using Dirac spinors $\Psi_i$, $l$ and $\nu$ and the $\gamma$ matrices are in the Dirac basis
 with $\gamma_5=i\gamma^0\gamma^1\gamma^2\gamma^3$ and
 $\sigma^{\mu\nu}=i(\gamma^\mu \gamma^\nu-\gamma^\nu \gamma^\mu)/2$. The
 cross-section for the two-particle process,  $l_1+N_2\rightarrow l_3+N_4$,
 where $l_1$ and $l_3$ are the initial and final state leptons, and $N_2$ and
 $N_4$ are the initial and final state nucleons, respectively, can be calculated
 from Fermi's Golden Rule. In the relativistic formalism, the differential cross-section for the process $1+2\rightarrow 3 +4$ is given by 
 \begin{equation}
 d\sigma =\frac{1}{(2E_1)(2E_2^*)~v_{\rm rel}}~\left\langle \left| M \right|^2
 \right \rangle~d\Phi_{\rm 34}
 (2 \pi)^4 \delta^4(p_1^\mu + p_2^\mu - p_3^\mu - p_4^\mu) \,,
 \end{equation}
 where $v_{\rm rel}$ is the relative velocity between particles in the initial state, 
 \begin{equation}
 d\Phi_{\rm 34}=\frac{d^3 p_3}{(2\pi)^3 2 E_3} 
\frac{d^3 p_4}{(2\pi)^3 2 E_4^*}(1-f_3) (1-f_4)\,,
 \end{equation}
is the Lorentz invariant phase which includes effects due to Pauli blocking of the final states and $\langle | M | ^2 \rangle$ is the square of the matrix element -- averaged over initial spin states
 and summed over the final spin states. Above, 
$E_i^*=\sqrt{p_i^2+(M^*_i)^2}$ and $M^*_i$ are the nucleon effective masses in
the medium. The differential absorption rate for a neutrino with energy $E_1$ in the medium where the density of the particle $2$ is $n_2$ is given by 
\begin{eqnarray}
\label{eq:sigma_FGR}
d \Gamma(E_1) = \langle n_2~v_{\rm rel}~d\sigma \rangle = 
2 \int \frac{d^3 p_2}{(2\pi)^3} ~f_2~v_{\rm rel}~d\sigma\,,
\end{eqnarray}
where $f_2$ is the distribution of the particle $2$ in the medium and the factor
of $2$ on the RHS accounts for its spin degeneracy. The distribution
functions $f_i$ are assumed to be Fermi-Dirac distributions characterized by
chemical potential $\mu_i$ and temperature $T$. 
Using the standard decomposition of the square of weak matrix element for free
nucleons in terms of the lepton tensor and the baryon tensor, we find that 
\begin{equation}
\left\langle \left| M \right|^2 \right \rangle = \frac{C^2 G_F^2}{4}
L_{\mu \nu} \Lambda^{\mu \nu}\,.
\end{equation}

The lepton tensor is  
\begin{eqnarray}
L^{\mu \nu} = \textrm{Tr}\left[
(-\fsl{p}_1 + m_1) \gamma^\mu(1-\gamma^5) 
(-\fsl{p}_3 + m_3) \gamma^\nu(1-\gamma^5)  
\right] \, ,
\label{eq:Lmunu}
\end{eqnarray}
where $q^\mu = p_1^\mu - p_3^\mu = p_4^\mu - p_2^\mu $ is the
energy-momentum transfer from the leptons to the baryons. In our case 
since particle 1 is a neutrino  $m_1\approx 0$ and 
$m_3=m_l$ where $m_l$ is the final charged lepton,  $m_l=m_e$ for electrons and $m_l=m_\mu$ muons, in the final state. 
The upper sign is for neutrinos while the lower sign is for antineutrinos, due to their left and right
handed character. We use the standard Feynman slash notation, where a slash denotes contraction of a four-vector with the
gamma matrices. 

Inspecting the kinematics of the leptons gives the allowed
range of values for the energy and momentum transfer to the nucleons for given four-momentum of
particle 1,
\begin{eqnarray}
q &=& \sqrt{p_1^2 + p_3^2 - 2 p_1 p_3 \mu_{13}} \\
q_0 &=& E_1 - E_3 \,,
\end{eqnarray}
where $\mu_{13}$ is the cosine of the angle between the momentum vectors of
particles one and three and $p_i$ is the magnitude of the momentum of particle
$i$. The maximum and minimum values of this
expression shows that the allowed range of momentum transfers to be $|p_1 - p_3|
< q < p_1 + p_3$. When both particles one and three are massless, these
relations imply $q_\mu^2 < 0$ and $|q_0| < q < 2E_1 - q_0$, but these
constraints do not hold for charged current reactions in which the final state lepton mass cannot be neglected.

The hadronic part of the matrix element is well known in the case of free
nucleons, and including mean field corrections in the nucleon spinors only slightly
alters its structure. The necessary modifications to the spin-sums are described
in Appendix \ref{app:MeanField}.
Then, the baryon contribution to the matrix element in the mean field
approximation is given by 
\begin{eqnarray}
\Lambda^{\mu \nu} = \textrm{Tr}\biggl[
(-\fsl{\tilde p}_2 + M^*_2) 
\left\{ g_V\gamma^\mu -g_A\gamma^\mu \gamma^5 
+ F_2\frac{i \sigma^{\mu \alpha} \tilde q_\alpha}{2 M_p} \right\}
~ (-\fsl{\tilde p}_4 + M^*_4)
\left \{ g_V\gamma^\nu -g_A\gamma^\nu \gamma^5 
- F_2\frac{i \sigma^{\nu \alpha} \tilde q_\alpha}{2 M_p} \right\}
\biggr] \,. 
\end{eqnarray}
Here $\tilde{p}^\mu_2=(E^*_2,\vec{p}_2)$, $\tilde{p}^\mu_4=(E^*_4,\vec{p}_4)$,
and
$\tilde q^\mu = \tilde p_4^\mu - \tilde p_2^\mu$. In the presence of background mean fields, the
nucleon energies $ E_2=E_k^*+U_2$ and $E_4=E_{k+q}^*+U_4$, where $U_2$ and $U_4$
are mean field potentials for 2 and 4, respectively. The effective masses of the
nucleons 2 and 4 in the medium are $M^*_2$ and $M^*_4$. 

We can now recast the absorption rate in Eq.~\ref{eq:sigma_FGR} as 
\begin{eqnarray}
\frac{d \Gamma(E_1)}{dE_3 d \mu_{13}} 
&=& \frac{C^2 G_F^2 \, p_3 (1-f_3(E_3))~L_{\mu \nu} ~{\cal I}^{\mu \nu}}
{32 \pi^2 E_1(1 - \exp{\left(-(q_0+\Delta \mu)/T\right)})} 
\label{eq:dsigFGR}
\end{eqnarray}
as in \cite{Reddy:99}. 
The nuclear part is now
factored and contained in the tensor  
\begin{eqnarray}
{\cal I}^{\mu \nu} &=& 
\int \frac{d^3p_2}{(2 \pi)^3 2 E_2^*}  
\int \frac{d^3p_4}{(2 \pi)^3 2 E_4^*}
 ~(f_2(E_2)-f_4(E_4))~\Lambda^{\mu \nu} (2\pi)^4 \delta^3(\vec{p}_2-\vec{p}_4-\vec{q})\delta(E_2-E_4-q_0)
\nonumber\\
&=\int& \frac{d^3 p_2}{(2 \pi)^2} ~\Lambda^{\mu \nu}~ 
\frac{f_2(E_2)-f_4(E_4)}{4 E_2^* E_4^*} \delta(E_2-E_4-q_0)\,,
\nonumber\\
\label{eq:Imunu}
\end{eqnarray}
where $E_{k+q}^*=\sqrt{(\vec{k}+\vec{q})^2+(M_4^*)^2}$ and in the second line we
have employed the momentum space Dirac delta function.  The detailed balance
factor for charged current reactions, $(1 - \exp{\left(-(q_0+\Delta
\mu)/T\right)})^{-1}$, comes from using the relation $f_2(1-f_4) = (f_2 - f_4)/(1 -
\exp[-(E_4 - E_2 - \mu_4 - \mu_2)/T])$.}

Eq.~\ref{eq:dsigFGR} together with Eq.~\ref{eq:Imunu} can be used to calculate
the charged current opacity. This would include corrections due to mean field
potentials, relativistic kinematics and weak magnetism. We calculate ${\cal
I}_{\mu \nu}$ in detail in section \ref{sec:practical}, but first we show that
the same result can be found from linear response theory.  

\subsection{The Charged Current Polarization Tensor}
\label{sec:polarization} 
The neutrino absorption rate in nuclear matter can be calculated using linear
response theory because at leading order in the weak interaction, the nucleonic
and leptonic parts factorize. For the weak interaction Lagrangian in
Eq.~\ref{eq:Lweak} linear response theory predicts \cite{Horowitz:91,Fetter:71}
\begin{equation}
\frac {d\Gamma(E_1)}{dE_3 d\mu_{13} } = \frac
{C^2 G_F^2}{32\pi^2} \frac{p_3}{E_1}
(1-f_3(E_3)) 
L_{\mu\nu}~{\cal S}^{\mu\nu}(q_0,q)\,,
\label{eq:LRdcross}
\end{equation}
where $L_{\mu\nu}$ is the lepton tensor defined earlier in Eq.~\ref{eq:Lmunu}, 
\begin{equation}
{\cal S}^{\mu\nu}(q_0,q)=\frac{-2~{\rm Im}~\Pol^{\mu\nu}(q_0,q)}{1 - \exp{\left(-(q_0+\Delta \mu)/T\right)}}\,,
\label{eq:dynamic}
\end{equation} 
 is called the dynamic response function, and  
 \begin{equation}
 \Pol^{\mu\nu}(q_0,q)= -i\int dt~d^3x~\theta(t)~e^{i(q_0t-\vec{q}\cdot \vec{x})}\langle ~|[j_\mu~(\vec{x},t),j_\nu(\vec{0},0)]|~\rangle\,,
\label{eq:correlation}
\end{equation}
is the retarded current-current correlation function or the polarization tensor where
$j_\mu$ is the weak charged current defined in Eq.~\ref{eq:Jweak} and $\langle |\cdots|\rangle$ is the thermodynamic average $={\rm Tr}~[{\rm exp}{(\beta(H-\sum_i\mu_i N_i))}\cdots]/{\cal Z}$ where ${\cal Z}$ is the grand canonical partition function. The
relationship in Eq.~ \ref{eq:dynamic} between the correlation function and the
dynamic structure factor is often called the {\it fluctuation-dissipation}
theorem \cite{Kubo:57, Landau:80}.   

This correlation function encodes all of the complexities associated with interaction between nucleons in the plasma and is in general difficult to calculate.     
When nucleons are treated as non-interacting particles, the polarization tensor can be calculated using the free single particle Greens functions.
We use the imaginary time
formalism \cite{Le-Bellac:00}, where the free nucleon propagator at zero chemical
potential is given by 
\begin{equation}
\script{G_F}(i \omega_n, {\bf{p}}) 
= \frac{M - \fsl{p}}{E_p^2 - (i \omega_n)^2}\,. 
\end{equation} 
where $\omega_n$ is a Fermionic Matsubara frequency. The extension to non-zero chemical potential is straightforward and is obtained by the replacement $i
\omega_n \rightarrow i \omega_n + \mu$ (see \cite{Le-Bellac:00} equation 5.70). The effects due to a space-time independent background 
mean field potential can also be similarly included since its contribution to the grand canonical Hamiltonian is proportional to $\int d^3x \bar \Psi \gamma^0
\Psi$, similar to the chemical potential (see Appendix~\ref{app:MeanField}). Additionally, the numerator, which comes from a spin sum, should be replaced by
the spin sums described in Appendix~\ref{app:MeanField}. 
These considerations imply that the propagator for nucleons in the dense medium is obtained by replacement $i \omega_n \rightarrow i \omega_n +
\nu_i$, where $\nu_i = \mu_i - U_i$, and $-\fsl{p} + M \rightarrow  -\fsl{\tilde p} +
M^*$ which gives 
\begin{equation}
\script{G}_{i,MF}(i \omega_n + \nu, {\bf{p}}) 
= \frac{M^* - \fsl{\tilde p}}{{E^*_{p,i}}^2 - (i \omega_n + \nu_i)^2},
\label{eq:G_MF}
\end{equation} 
where $M^*$ is the effective mass, $\tilde p^\mu = (\pm E^*_{i,p}, -\vec{p})$, and
$E^*_{i,p} = \sqrt{p^2+(M_i^*)^2}$.   
Using these propagators, the imaginary time polarization functions are given by
\begin{widetext}
\begin{fmffile}{polarization}
\unitlength = 0.7mm
\begin{eqnarray}
\Pol_{ab}(i\omega_m - \Delta \mu, \vec{q}) =
&&
\begin{gathered}
\begin{fmfgraph*}(80,40)
  \fmfleft{i0}
  \fmfright{o2}
  \fmf{wiggly,label=$i\omega_m - \Delta \mu,, \vec{q}$}{i0,i1}
  \fmf{wiggly,label=$i\omega_m - \Delta \mu,, \vec{q}$}{o1,o2}
  \fmf{plain_arrow,left,tension=0.7,label=$i(\omega_n + \omega_m) + \nu_4,,
  \vec{k}+\vec{q}$}{i1,o1}
  \fmf{plain_arrow,left,tension=0.7,label=$i\omega_n + \nu_2,, \vec{k}$}{o1,i1}
  \fmfv{decor.shape=circle,decor.filled=empty,decor.size=10,label=$\Gamma_a$,label.a=0}{i1}
  \fmfv{decor.shape=square,decor.filled=empty,decor.size=10,label=$\Gamma_b$,label.a=180}{o1}
  \fmflabel{$\Gamma_a$}{i1}
\end{fmfgraph*}
\end{gathered}
\qquad \nonumber\\ &&
= {T} \sum_n \int \frac{d^3 k}{(2 \pi)^3} 
{\rm{Tr}} \left[\script{G}_{2,MF}(i \omega_n + \nu_2, {\vec{k}}) \Gamma_a
  \script{G}_{4,MF}(i \omega_m + i \omega_n + \nu_4, {\vec{k}}+{\vec{q}}) \Gamma_b
  \right],
\end{eqnarray}
\end{fmffile}
\end{widetext}
where the $\Gamma_a$ represent different interaction vertices (i.e. $C_V
\gamma^\mu$, etc.), $i \omega_m$ is a Bosonic Matsubara frequency, and $\Delta
\mu = \mu_2 - \mu_4$. 

Using the Matsubara sum results from appendix~\ref{app:Matsubara} and the baryon
tensor portion of the weak interaction matrix element given above, we find that the imaginary
time polarization tensor is 
\begin{widetext}
\begin{eqnarray}
\Pol_{\mu \nu}(i \omega_m - \Delta \nu, \vec{q}) 
&=& \int \frac{d^3 k}{(2 \pi)^3} \frac{\Lambda_{\mu \nu}}{4 E_{2,k}^* E_{4,k+q}^*}
\biggl[\frac{f_2(E_{2,k}^*) - f_4(E_{4,k+q}^*)}{i \omega_m - \Delta \nu +
E_{2,k}^* - E_{4,k+q}^*} 
-\frac{\bar f_2(E_{2,k}^*) - \bar f_4(E_{4,k+q}^*)}{i \omega_m - \Delta \nu -
E_{2,k}^* + E_{4,k+q}^*} 
\nonumber\\&& 
+ \frac{1 - f_2(E_{2,k}^*) - \bar f_4(E_{4,k+q}^*)}{i \omega_m - \Delta \nu +
E_{2,k}^* + E_{4,k+q}^*}
- \frac{1 - \bar f_2(E_{2,k}^*) - f_4(E_{4,k+q}^*)}{i \omega_m - \Delta \nu -
  E_{2,k}^* - E_{4,k+q}^*}
\biggr],
\label{eq:PolFnc}
\end{eqnarray}
\end{widetext}
where $\Delta \nu = \nu_2 - \nu_4$.

When considering scattering and capture processes, we only require portions of
the polarization that are non-zero for $\tilde q_\alpha^2 < (M_2^*-M_4^*)^2$ due to kinematic
restrictions, so the last two pieces of the polarization can be ignored. The
piece of the polarization containing only anti-baryon distribution functions is only non-zero at very
high temperatures not encountered in supernovae and neutron star mergers.
Therefore, the last three terms in Eq.~\ref{eq:PolFnc}
will be neglected in the rest of the discussion, although we keep in mind the
second term will be present when calculating scattering rates from electrons and
positrons.

For the linear response, we require the real-time polarization function. This
can be found by analytically continuing the imaginary-time polarization
\cite{Fetter:71} via the replacement $i \omega_m - \Delta \mu \rightarrow q_0
+ i \eta$. The need for this particular replacement can be seen in the incoming
and outgoing Bosonic lines in the bubble diagram above, which include a chemical
potential difference because the Bosonic frequencies carry isospin charge. 
Analytically continuing to real-time and using the relation 
\begin{equation}
\frac{1}{\omega \pm i \eta} = \script{P} \frac{1}{\omega} 
\mp i \pi \delta(\omega),
\end{equation}
we then find the real and imaginary parts of the polarization tensor are given
by  
\begin{widetext}
\begin{eqnarray}
\label{eq:impol}
\im{\Pol}^R_{\mu\nu}(q_0,q) &=& -\pi \int  \frac{d^3k \,
\Lambda_{\mu\nu}}{(2\pi)^3 4 E_{2,k}^* E_{4,k+q}^*}  
\delta(E_{2,k}^* - E_{4,k+q}^* + \tilde q_0) \{f_2(E_{2,k}^*) - f_4(E_{4,k+q}^*)\}
\\
\re{\Pol}^R_{\mu\nu}(q_0, q) &=& 
\script{P} \int \frac{d^3 k \, \Lambda_{\mu\nu}}{(2 \pi)^3 4 E^*_{2,k} E^*_{4,k+q}}  
\frac{f_2(E_{2,k}^*) - f_4(E_{4,k+q}^*)}{E^*_{2,k} + \tilde q_0 - E^*_{4,k+q}},
\end{eqnarray}
\end{widetext}
where $\tilde q_0 = q_0 + U_2 - U_4$.
Note that we are not taking the imaginary part of $\Lambda$ in this expression.

From this, it is clear that the mean field polarization function is just the 
free fermion polarization function with the replacements 
\begin{eqnarray}
\mu_i &\rightarrow& \nu_i \\
q^\mu &\rightarrow& \tilde q^\mu = (q_0 + U_2 - U_4, -\vec{q}),
\end{eqnarray} 
where $\tilde q^\mu$ is the kinetic energy and momentum transfer from the
entrance channel nucleon to the final state nucleon. 
Comparing Eq.~\ref{eq:dsigFGR} and Eq.~\ref{eq:LRdcross} we see that linear
response theory and the Fermi's Golden Rule approach would yield the same result when 
\begin{equation}
{\cal I}^{\mu \nu}(q_0,q) =(1 - \exp{\left(-(q_0+\Delta \mu)/T \right)}){\cal S}^{\mu\nu}(q_0,q)\,. 
\end{equation} 
Neglecting anti-particle and pair contributions in Eq.~\ref{eq:impol}, we can verify the above equation
to prove that these approaches are equivalent when correlations are neglected. 
As we will find, projecting ${\cal I}_{\mu \nu}$ along and orthogonal to
$\tilde q^\mu$ results in simple expressions for the polarizations analogous to
the results found in the literature for the case of neutral current polarization
tensors. 

The advantage of the linear response formalism is that it can be extended to included higher
order corrections to the medium response, which cannot be done systematically
using the Fermi's Golden Rule approach.  For instance, the approach above
neglects effects due to screening of the weak interaction by particle-hole pairs
in the medium and collective excitations such as the giant isovector dipole resonance and the Gammow-Teller resonance, both of which arise due to strong
interactions between nucleons. To include these effects consistently with the
mean field ground state of the nuclear medium, the response should be calculated
in the Random Phase Approximation (RPA) as discussed earlier in \cite{Reddy:99}.
The response functions in RPA can be formulated using the real and imaginary parts of the polarization tensors derived
here. 

\subsection{Practical Expressions for $\cal{I}^{\mu \nu}$ and $L^{\mu\nu}$} 
\label{sec:practical} 

The results described in preceding sections provide formulae to calculate  the charged current absorption rates, including weak magnetism and mean field contributions, but
their forms are not amenable for use in numerical simulations.
Here, we derive simple expressions for the components of the
differential absorption rate that can easily be implemented for practical
calculations.

\subsubsection{Expressions for ${\cal I}^{\mu\nu}$}
\label{sec:full_response}
First we consider the general for of the integrals given in equation
\ref{eq:Imunu}.
We employ the energy space delta function in Eq.~\ref{eq:Imunu} to
remove the integrals over the nucleon angle and leave a single integral over
energy. 
Transforming the energy space delta function to a delta function in the cosine
of the angle between nucleons and enforcing momentum conservation (with the
momentum transfered assumed to be in the z-direction), we can write
structure function as 
\begin{eqnarray}
{\cal I}^{\mu \nu} = \frac{1}{(4 \pi)^2 q} \int_{M_2^*}^\infty dE_2 
\int d \Omega_2 \delta(\mu - \mu_0) 
\theta(E_2-e_{m})
(f_2 - f_4) \Lambda^{\mu \nu},
\end{eqnarray}
with $\mu_0 = (\tilde q_\mu^2 + 2 E^*_2 \tilde q_0 + 
{M_2^*}^2 - {M_4^*}^2)/2 p_2 q$.
Here, $\beta= 1 + (M_{*,2}^2 - M_{*,4}^2)/\tilde q_\mu^2$ and
\begin{equation}
e_{m}=-\beta \frac{\tilde q_0}{2} 
+ \frac{q}{2} \sqrt{\beta^2 - \frac{4 {M_2^*}^2}{\tilde q_\mu^2}}.
\label{eq:eminus1}
\end{equation}
The physical meaning of this lower limit becomes clearer when it is expressed in
terms of \cite{Leinson:01}
\begin{equation}
\sigma_{\pm} = 1 - (M^*_2 \pm M^*_4)^2/\tilde q_\alpha^2.
\end{equation} 
The result is 
\begin{equation}
e_m = -\beta \frac{\tilde q_0}{2} + \frac{q}{2} \sqrt{\sigma_+ \sigma_-}. 
\label{eq:eminus2}
\end{equation} 
The allowed range of $\tilde q_\alpha^2$ is then given by the range of values
for which $\sigma_+ \sigma_- \geq 0$, which correspond to $\tilde q_\alpha^2 <
(M_2^* - M_4^*)^2$ or $\tilde q_\alpha^2 > (M_2^* + M_4^*)^2$. Clearly, the
first condition enforces the impact of the mass difference in capture processes
while the second condition is the appropriate kinematic condition for pair
production.  

The only terms in the integrated baryon tensor ${\cal I}^{\mu \nu}$ that cannot
be pulled outside of the integral are power of $p_2^\mu$, since $p_4^\mu =
p_2^\mu + \tilde q^\mu$.  Therefore, ${\cal I}^{\mu \nu}$ can be expressed in
terms of the tensors
\begin{equation}
\pp^{\{a_k\}} = \frac{8}{(4\pi)^2 q} \int_{e_m}^\infty dE_2 \int d\Omega_2 
\Theta(\mu-\mu_0) (f_2 - f_4) \tilde p_2^{a_1} ... \tilde p_2^{a_k}.
\end{equation}
We include the extra factor of 8 for convenience, since
$\textrm{Tr}[\Lambda^{\mu \nu}]$ contains a factor of 8.

Since the nucleons are on-shell, it is easy to show that 
\begin{equation}
\label{eq:pol_projection}
\tilde q_\mu \pp^{\mu \{a_k\}} = - \frac{\beta \tilde q_\mu^2}{2}\pp^{ \{a_k\}}.
\end{equation}
This relation comes in handy when trying to simplify the different parts of the 
polarization; when $\Lambda^{\mu \nu}$ contains $(p_2 \cdot q)$, it can be
replaced with $-\beta q_\mu^2/2$.  

The expression for $I^{\mu \nu}$  simplifies if decompose it 
relative to $\tilde q^\mu$. The set of projection tensors that describe
this decomposition are shown in Appendix~\ref{app:Decomposition}. This results in the
expansion
\begin{equation}
\pp_{\mu \nu} = \beta^2 \pp_Q P^Q_{\mu \nu} 
+ \pp_L P^L_{\mu \nu}
+ \pp_T P^{T+}_{\mu \nu}
+ \beta \pp_{M} P^{M+}_{\mu \nu},  
\end{equation} 
where we have pulled out factors of $\beta$ for convenience. The explicit form
of these expansion terms are 
\begin{eqnarray}
\pp_Q &=& \frac{\tilde q_\mu^2}{4 \pi q} \int_{e_m}^\infty dE_2^*(f_2 - f_4) \\
\label{eq:IsQ}
\pp_L &=&-\frac{\tilde q_\mu^2}{4 \pi q^3} \int_{e_m}^\infty dE_2^* (f_2 - f_4) 
(2 E_2^* + \beta \tilde q_0)^2 \\
\label{eq:IsL}
\pp_M &=&-\frac{\tilde q_\mu^2}{4 \pi q^2} \int_{e_m}^\infty dE_2^* (f_2 - f_4) 
(2 E_2^* + \beta \tilde q_0) \\
\label{eq:IsM}
\pp_T &=& - \frac{1}{2}\pp_L 
+ \left(\frac{2 m_2^2}{\tilde q_\alpha^2} - \frac{\beta^2}{2}\right)\pp_Q.
\label{eq:IsT}
\end{eqnarray}
Using Eq.~\ref{eq:pol_projection}, we then find for the lower rank tensors 
\begin{eqnarray}
\pp^\nu &=& -\frac{2 \tilde q^\nu \beta }{\tilde q_\alpha^2} \pp_Q 
- \frac{2 n^\nu}{\tilde q_\alpha^2} \pp_M \\
\pp &=& \frac{4 }{\tilde q_\alpha^2} \pp_Q. 
\end{eqnarray}

We are now left with one dimensional integrals over $E_2$ that can be 
expressed in terms of ultra-relativistic Fermi integrals. 
We then find the basic pieces of ${\cal I}^{\mu\nu}$ are given by 
\begin{eqnarray}
\pp_Q &=& \frac{\tilde q_\mu^2 T}{4 \pi q}  \Gamma_0 \\
\pp_L &=& -\frac{\tilde q_\mu^2 T^3}{4 \pi q^3}  
\left[a^2 \Gamma_0 + 4 a \Gamma_1 
+ 4 \Gamma_2 \right] \\
\pp_M &=& -\frac{\tilde q_\mu^2 T^2}{4 \pi q^2}  
\left[a \Gamma_0
+ 2 \Gamma_1 \right] \\
\pp_T &=& - \frac{1}{2}\pp_L 
+ \left(\frac{2 M_{*,2}^2}{\tilde q_\alpha^2} - \frac{\beta^2}{2}\right)\pp_Q
\end{eqnarray}
where
\begin{eqnarray}
a &=& (\beta \tilde q_0/T + 2e_m/T) \\
\delta_1 &=& (\mu_2 - U_2 - e_m)/T \\
\delta_2 &=&  (\mu_4 - U_4 - \tilde q_0 - e_m)/T.
\end{eqnarray}
with
\begin{eqnarray}
\Gamma_n(\delta_2, \delta_4) =
\int_0^\infty dx ~x^n ~ \left(f_{\rm FD}(x-\delta_2) - f_{\rm
FD}(x-\delta_4)\right)\, ,
\end{eqnarray}  
where $f_{FD}(x) = 1/[\exp(x) + 1]$.
The function $\Gamma_0 = \ln [(\exp \delta_2+1)/(\exp \delta_4+1)]$ has a
simple analytic form, while the other $\Gamma_n$ are related to polylogarithmic
functions and can either be tabulated or calculated using the highly accurate
approximations given in \cite{Takahashi:78}.

All that is left to do is evaluate the trace in $\Lambda_{\mu \nu}$ and then
decompose the resulting imaginary part of the real-time polarization tensor
using the tensors described in Appendix~\ref{app:Decomposition}. 
We choose to split the baryon tensor into coefficients of the various
combinations of weak coupling constants, which defines the quantities 
\begin{eqnarray}
\label{eq:Imunu_decomp}
{\cal I}_{\mu \nu} =
 g_V^2    \pp^{V }_{\mu \nu} 
+ g_A^2    \pp^{A }_{\mu \nu} 
+ F_2^2    \pp^{T }_{\mu \nu} + g_V g_A  \pp^{VA}_{\mu \nu} 
+ g_V F_2  \pp^{VT}_{\mu \nu} 
+ g_A F_2  \pp^{AT}_{\mu \nu}.
\end{eqnarray} 
Each of these coefficient tensors are then decomposed in the form 
\begin{eqnarray}
\pp^{i,\mu \nu} = \pp^i_Q P_Q^{\mu \nu} + \pp^i_L P_L^{\mu \nu} 
+ \pp^i_{T+} P_{T+}^{\mu \nu} + \pp^i_{T-} P_{T-}^{\mu \nu}
+ \pp^i_{M+} P_{M+}^{\mu \nu} + \pp^i_{M-} P_{M-}^{\mu \nu},
\end{eqnarray} 
where $i$ denotes either the vector, axial, tensor or mixed portions of the weak
interaction.

Calculating the trace of $\Lambda_{\mu\nu}$ is straightforward (we employ the
Mathematica package \cite{Mertig:91}). 
Using $\Lambda_{\mu\nu}$ in equation~\ref{eq:impol}, we find  
the non-zero components of the vector polarization are 
\begin{eqnarray}
\pp_Q^{V} &=&  
\left(\lambda^2 + \sigma_- - 1\right) \pp_Q \\ 
\pp_L^{V} &=& \pp_L + \sigma_- \pp_Q  \\
\pp_{T+}^{V} &=& \pp_T + \sigma_- \pp_Q \\
\pp_{M+}^{V} &=& \pp_M \lambda. 
\end{eqnarray}
Here, we have defined $\lambda = \beta - 1$ and 
$\Delta = (M^*_4 - M^*_2)/M^*_2$.

The non-zero pieces of the axial polarization are
\begin{eqnarray}
\pp_Q^{A} &=& 
\left(\lambda^2 + \sigma_+ - 1\right) \pp_Q \\ 
\pp_L^{A} &=& \pp_L + \sigma_+ \pp_Q  \\
\pp_{T+}^{A} &=& \pp_T + \sigma_+ \pp_Q  \\
\pp_{M+}^{A} &=& \pp_M \lambda.
\end{eqnarray}

The non-zero pieces of the tensor polarization are
\begin{eqnarray}
\pp_L^{T} &=& \frac{ q_\alpha^2}{4m_2^2} 
\left[\left(\sigma_- - \beta^2 + \frac{4 m_2^2}{q_\alpha^2} \right)\pp_Q - \pp_L \right]
\\
\pp_{T+}^{T} &=& \frac{ q_\alpha^2}{4m_2^2} 
\left[\left(\sigma_- - \beta^2 + \frac{4 m_2^2}{q_\alpha^2} \right)\pp_Q - \pp_T
\right].
\end{eqnarray}
The mixed vector-tensor polarization is 
\begin{eqnarray}
\pp_L^{VT} &=&  (2 + \Delta \beta) \pp_Q \\
\pp_{T+}^{VT} &=&  (2 + \Delta \beta) \pp_Q \\
\pp_{M+}^{VT} &=& -\frac{ \Delta}{2} \pp_M .
\end{eqnarray}
The vector axial piece is 
\begin{equation}
\pp^{VA}_{T-} = - i 2 \pp_M. 
\end{equation}
And finally the axial tensor piece is
\begin{equation}
\pp^{AT}_{T-} = - i(2+\Delta) \pp_M.
\end{equation}
These expressions are similar to those found in \cite{Reddy:98, Horowitz:03},
but they include extra terms depending on the mass and potential differences of the nucleons.
A detailed comparison between these results and previous results is made in
section \ref{sec:comparison}.

\subsubsection{The Lepton Tensor} 
\label{sec:lepton}
The contraction of the lepton tensor with the imaginary part of the polarization
is the last piece required to calculate the differential neutrino cross-section.
The easiest way to perform this contraction is to project the lepton tensor
using the same projectors we used for the polarization tensor. 
Performing these projections gives 
\begin{eqnarray}
L_{L} &= \frac{8}{\tilde q_\alpha^2} [&
-2(\tilde n \cdot p_1)^2 
+ 2 (\tilde n \cdot p_1)(\tilde n \cdot q) 
+ \tilde q_\alpha^2 (p_1 \cdot q)]
\nonumber\\
L_{Q} &= \frac{8}{\tilde q_\alpha^2}[&
2(\tilde q \cdot p_1)^2 
- 2 (\tilde q \cdot p_1)(q \cdot \tilde q) 
+ \tilde q_\alpha^2 (p_1 \cdot q)]
\nonumber\\
L_{M+} &= \frac{8}{\tilde q_\alpha^2} [&
(\tilde q \cdot p_1)(\tilde n \cdot q) 
+ (\tilde n \cdot p_1)(q \cdot \tilde q - 2 \tilde q \cdot p_1) 
]
\nonumber\\
L_{M-} &= \frac{8 i}{\tilde q_\alpha^2} &\epsilon^{\alpha \beta \gamma \delta}
\tilde n_\alpha p_{1,\beta} \tilde q_\gamma q_\delta  
\nonumber\\
L_{T+} &= \frac{8}{\tilde q_\alpha^2} \bigl[ &(\tilde n \cdot p_1)^2 
- (\tilde n \cdot p_1)(n \cdot q) + (\tilde q \cdot p_1)(\tilde q \cdot q - p_1 \cdot \tilde q) \bigr]  
\nonumber\\
L_{T-} &= \frac{8 i}{\tilde q_\alpha^2} [& (\tilde n \cdot p_1)(\tilde q \cdot q)
- (\tilde n \cdot q) (p_1 \cdot \tilde q) ].
\end{eqnarray} 

The contractions appearing above are 
\begin{eqnarray}
p_1 \cdot q &=& \frac{q_\alpha^2-m_3^2}{2} \nonumber\\
p_1 \cdot \tilde q &=& p_1 \cdot q + \Delta U E_1 \nonumber\\
p_1 \cdot \tilde n &=& - \frac{q_\alpha^2}{2 q} [E_1 + E_3 - \Delta U + 2 \Delta U E_1 q_0/q_\alpha^2 + (q_0 - \Delta U) m_3^2/q_\alpha^2 ] 
\nonumber\\ 
\tilde n \cdot q &=& - q \Delta U 
\nonumber\\ 
q \cdot \tilde q &=& q_\alpha^2 + q_0 \Delta U.
\end{eqnarray}
In general, the projections of the lepton tensor have relatively complicated
forms since we are projecting the lepton tensor relative to $\tilde q_\mu$
rather than the more natural $q_\mu$. In the free gas limit and neglecting
$m_3$, these expressions reduce to 
\begin{eqnarray}
L_T &=& 8 q_\alpha^2 \left(A + 1\right) \\  
L_L &=&-8 q_\alpha^2 A \\
L_Q &=& 0 \\
L_{M+} &=& 0 \\
L_{T-} &=& \pm i 16 (n \cdot p_1),
\end{eqnarray}
where 
\begin{equation}
A = \frac{4 E_1 E_3 + q_\alpha^2}{2 q^2} = \frac{E_1 E_3}{q^2} (1 + \mu_{13}),
\end{equation}
which agrees with expressions previously found in the literature
\cite{Horowitz:03, Reddy:98}.

With these results, the contraction of the lepton tensor and the polarization 
tensor is quite simple (using the result of Appendix~\ref{app:Decomposition})
\begin{eqnarray}
L^{\mu \nu} {\cal I}_{\mu \nu} = \sum_{i=\{V,A,T,VA,VT,AT\}} {\cal C}_i\left[L_L \pp^i_L + L_Q \pp^i_Q - 2 L_{M+} \pp^i_{M+} 
- 2 L_{M-} \pp^i_{M-} 
+ 2 L_{T+} \pp^i_{T+} + 2 L_{T-} \pp^i_{T-} \right],
\end{eqnarray}
where ${\cal C}_i = \{g_V^2, g_A^2, F_2^2, g_A g_V,\ldots \}$ and the differential neutrino cross section can be found using equations
\ref{eq:dsigFGR} and \ref{eq:Imunu_decomp}.
 
\subsection{Opacities from a Limiting Form of the Matrix Element} 
\label{sec:approximations} 

Often, it is assumed that the nucleon masses dominate all other energy scales
entering the averaged matrix element \cite{Roberts:12b}. In this approximation,
\begin{eqnarray}
\langle | \script{M}|^2 \rangle &=& 16 C^2 G_F^2
\left[(g_V + g_A)^2 (p_1 \cdot \tilde p_2) (p_3 \cdot \tilde p_4) + (g_V - g_A)^2 (\tilde p_2 \cdot p_3) (p_1 \cdot \tilde p_4) - (g_V^2 - g_A^2) (p_1 \cdot p_3) (\tilde p_2 \cdot \tilde p_4) \right]
\nonumber\\ 
&\approx& 16 C^2 G_F^2 E_1 E_3 M^*_2 M^*_4
\left[(g_V + g_A)^2 + (g_V - g_A)^2- (g_V^2 - g_A^2) (1-\mu_{13})\right]\,,
\end{eqnarray}
where the second line is to leading order in the nucleon mass \citep{Giunti:07}. Following a
similar set of steps to those for the full matrix element, we find the
approximate cross-section is given by 
\begin{eqnarray}
\label{eq:csec_constant_M}
\frac{d \Gamma(E_1)}{dE_3 d \mu_{13}} &\approx& \frac{C^2 G_F^2}{4 \pi^2} 
\frac{p_3 E_3}{1 - \exp[-(q_0 + \Delta \mu)/T]} 
\left[ g_V^2 (1+\mu_{13}) + g_A^2(3-\mu_{13})\right] \frac{4 M_2^*
M_4^*}{\tilde q_\alpha^2} I_Q.
\end{eqnarray}
This is similar in form to the non-relativistic cross-section given in
\cite{Reddy:98, Reddy:99}, except that non-relativistic
kinematics has not been assumed for the integrals over the nucleon momentum or
in $e_m$. When non-relativistic kinematics are enforced at all momenta, an upper
limit in the nucleon energy integral is required to enforce momentum
conservation which is clearly spurious. This upper limit gives the second
logarithmic factor shown in Eq. 48 of \cite{Reddy:99}.  Additionally, the
form of $e_m$ is different when it is calculated using non-relativistic
kinematics.

A further approximation that is often assumed due to the large mass of the
nucleons is $\vec{q}=0$ in the nucleon integrals and  $E_{2,4}^*=M_{2,4}^*$
\cite{Bruenn:85, Martinez-Pinedo:12}. In
this case, the response function of the medium becomes 
\begin{eqnarray}
I_{q=0} &=& 2 \pi \delta (E_2 - E_4 + q_0) 
\int \frac{d^3 p_2}{(2\pi)^3} (f_2 - f_4) \nonumber\\ 
&\approx& \pi \delta(M^*_2 + U_2  - M^*_4 - U_4 + q_0) (n_2 - n_4).  
\end{eqnarray} 
The second line follows from assuming $p^2/(2M^*_2) = p^2/(2M^*_4)$. Then, the
differential cross section is given by 
\begin{eqnarray}
\label{eq:csec_qeq0}
\frac{d \Gamma(E_1)}{dE_3 d \mu_{13}} &\approx& \frac{C^2 G_F^2}{2 \pi^2} 
\frac{p_3 E_3}{1 - \exp[(\Delta M^* + \Delta U - \Delta \mu)/T]}I_{q=0} \left[ g_V^2 (1+\mu_{13}) + g_A^2(3-\mu_{13})\right],
\end{eqnarray}
which gives the standard integrated cross section per volume 
\begin{equation}
\frac{\sigma}{V} = \frac{C^2 G_F^2}{\pi} (g_V^2 + 3g_A^2) p_3 E_3
\frac{n_2 - n_4}{1-\exp[(\Delta M^* + \Delta U - \Delta \mu)/T]}. 
\end{equation} 

These
approximate forms of the cross section will be compared to the full results
below.  

\section{Results} 
\label{sec:results} 

Here, we present differential cross sections and comparison to previous results
in the literature.
In the charged current rates, the potential energy difference between neutrons
and protons can play a dominant role in the capture mean free paths of neutrinos
in the medium.  This potential difference depends on the effective nuclear
interaction that is assumed and is therefore model dependent. For illustrative
purposes in this section, we choose a very simple density dependent potential energy given by 
\begin{equation}
\Delta U = \Delta U_0 \frac{(n_2 - n_4)}{n_{\rm sat}}\,,
\end{equation}
where $n_{\rm sat}=0.16$ fm$^{-3}$ and we choose a coupling value of $\Delta U_0 = 40
~\textrm{MeV}$ and keep the nucleon masses fixed at their vacuum values for all
densities. At very low and high densities, a linear model for the potential
energy is unrealistic. Therefore, we emphasize that the results using this model
for the potential are only for illustrative purposes. We choose this
model only for its simplicity and because it prevents us from having to choose
from the plethora of relativistic mean field theories available. Additionally,
we keep the nucleon masses fixed to their vacuum values for most of the
discussion for simplicity, although we consider the impact of varying the
effective nucleon mass at saturation density in section \ref{sec:eff_mass}. It is important to note that the 
formalism presented above fully accounts for the impact of effective nucleon
masses on the neutrino absorption rate and the publicly available code
associated with this paper can be used to assess the impact of any variation in
the nucleon effective masses.

\begin{figure*}[t] 
\includegraphics[width=0.49\textwidth]{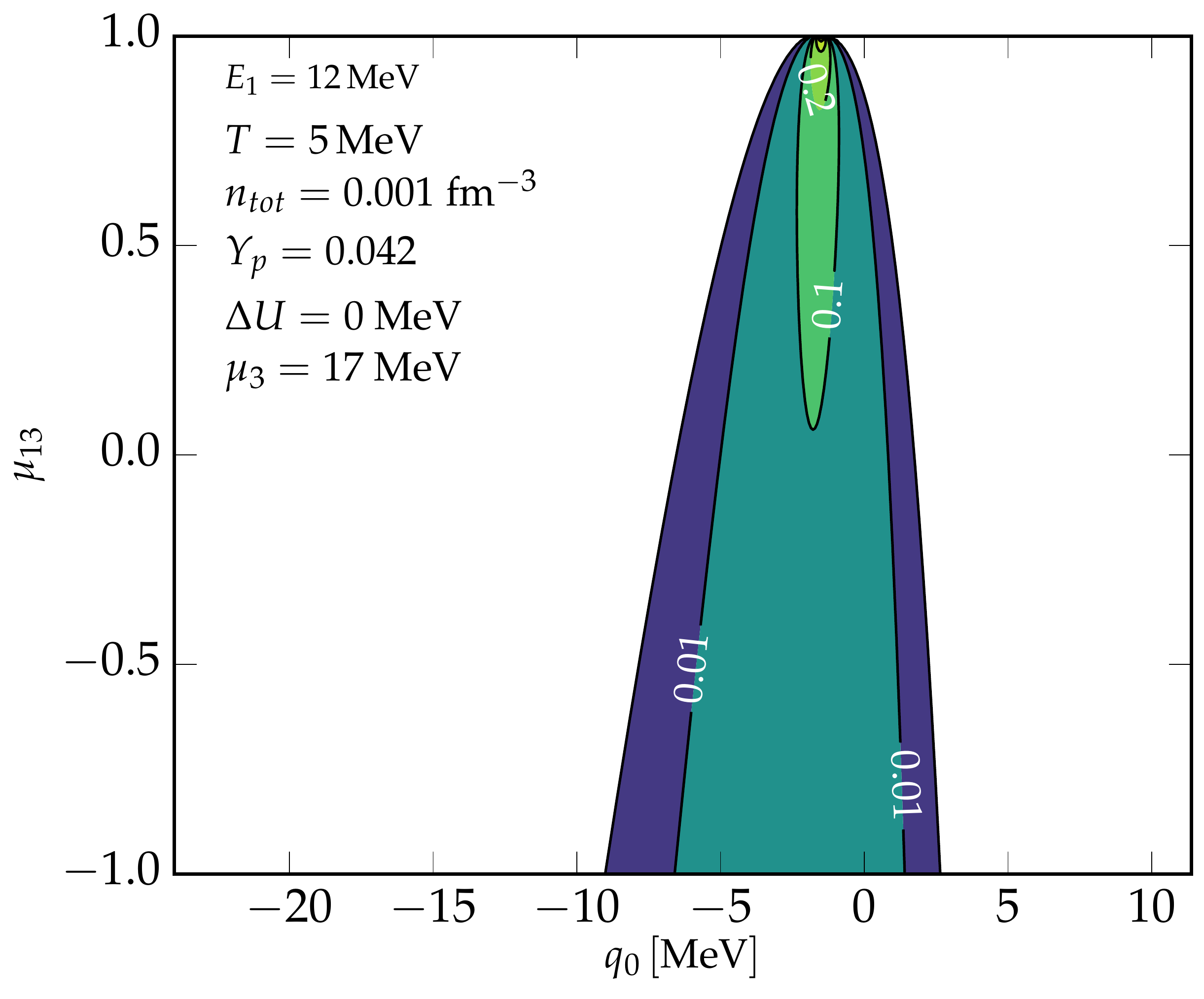} 
\includegraphics[width=0.49\textwidth]{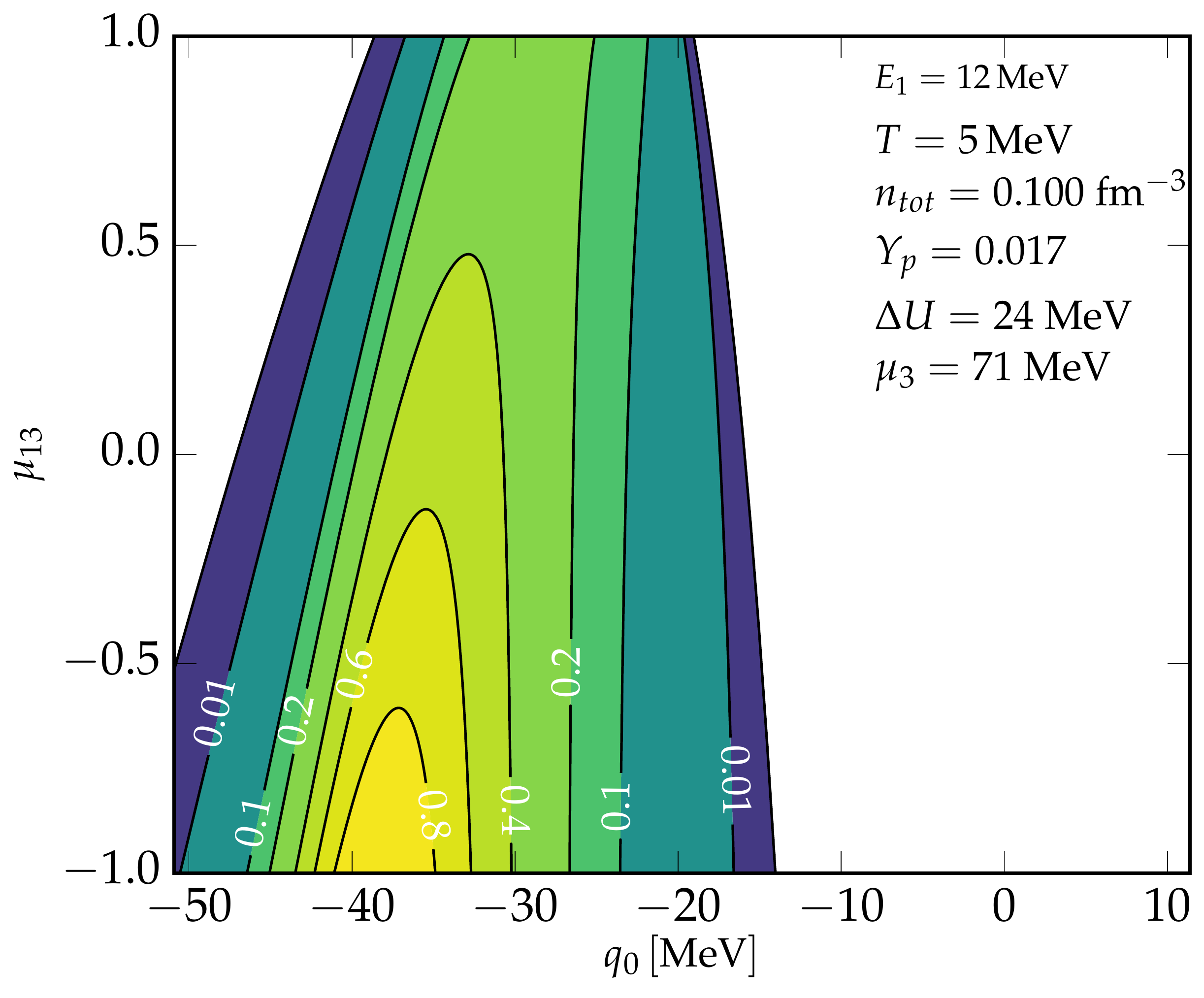} 
\caption{ Contours of the $\nu_e + n \rightarrow e^- + p$ double-differential absorption rate as
a function of electron scattering angle $\mu_{13}$ and energy transfer $q_0$,
normalized to the integrated absorption rate.}
\label{fig:dsigdq0dmu}
\end{figure*}

\begin{figure*}[t] 
\includegraphics[width=0.49\textwidth]{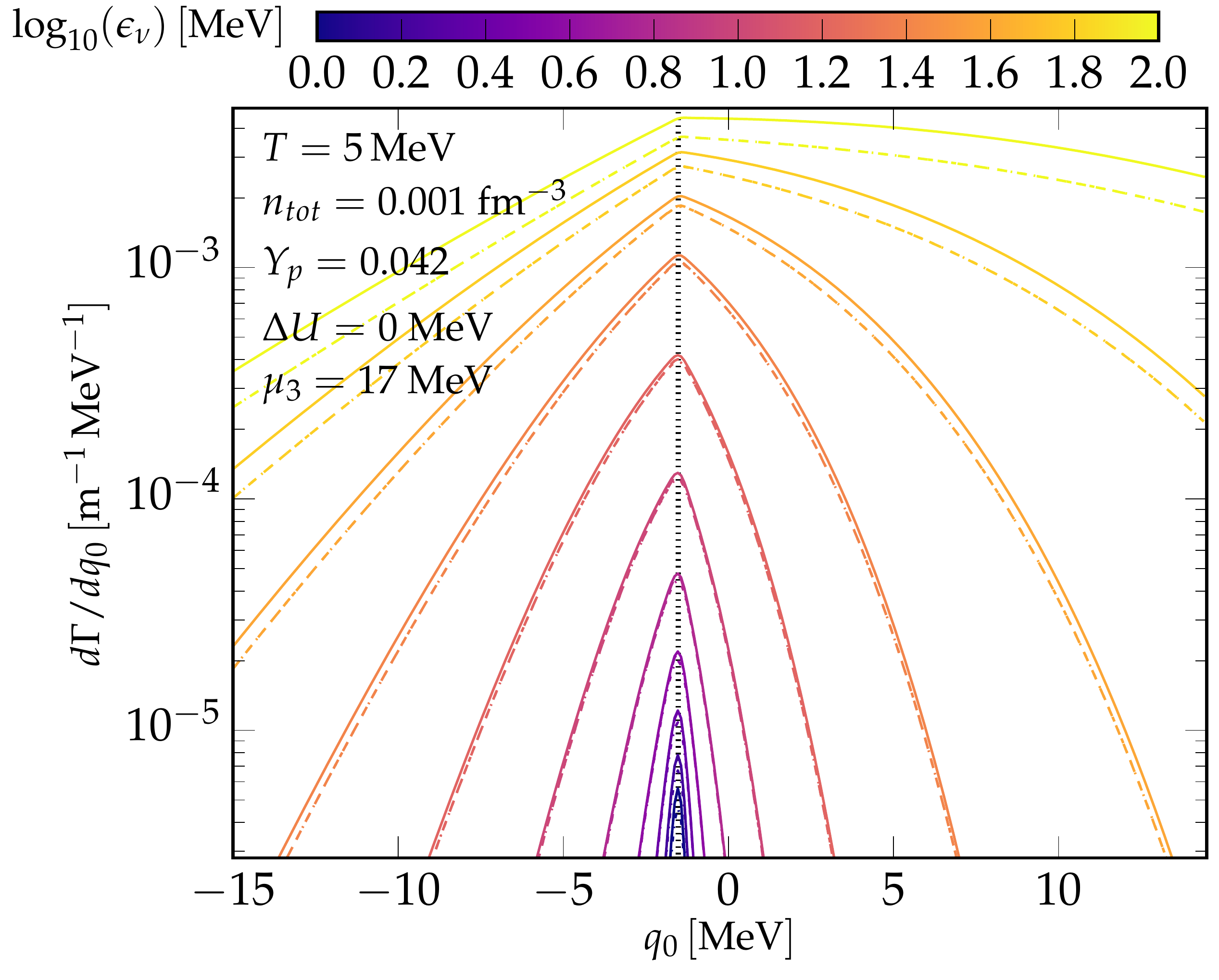} 
\includegraphics[width=0.49\textwidth]{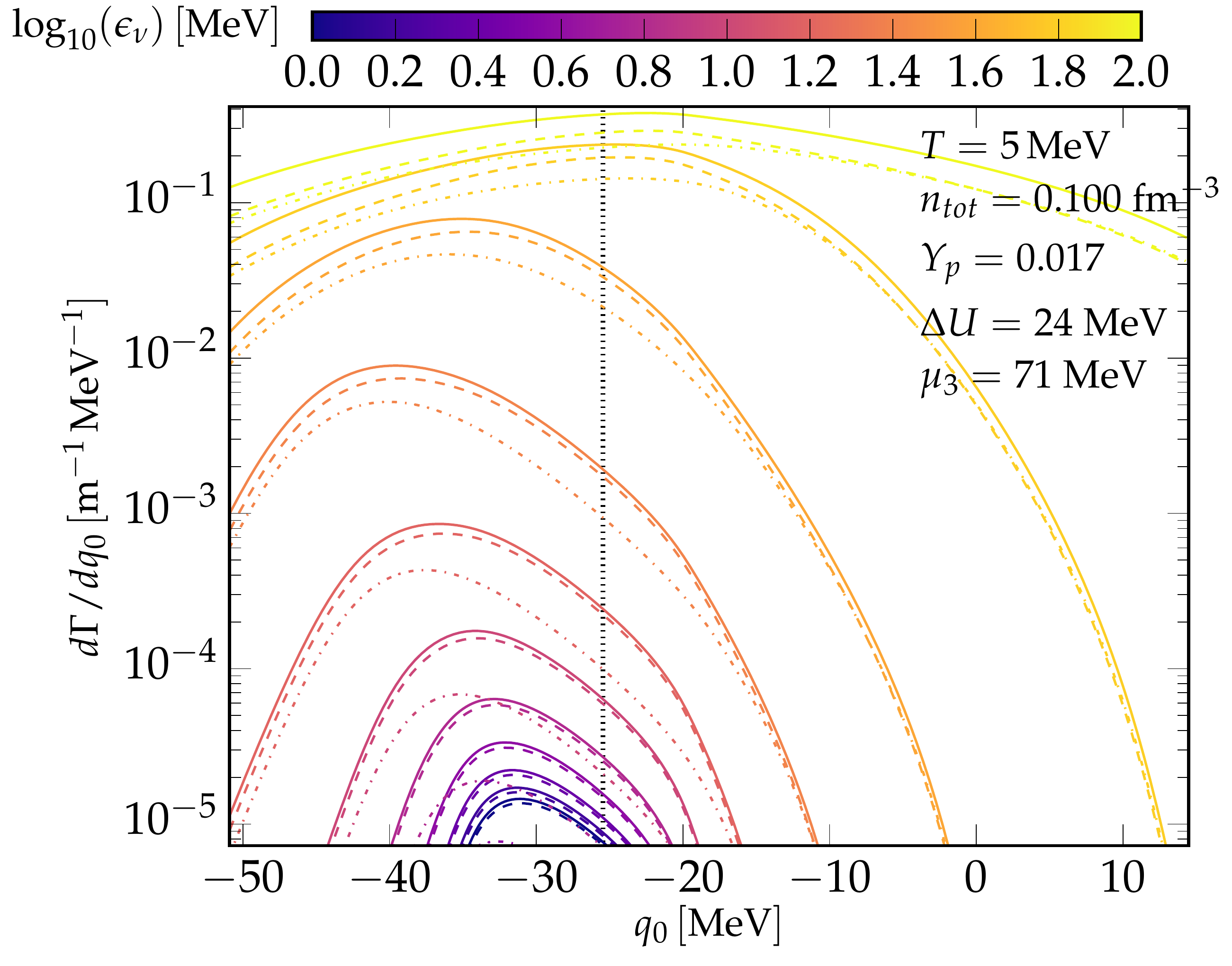} 
\caption{The angle integrated cross section as a function of energy transfer
$q_0$ for $\nu_e + n \rightarrow e^- + p$.  
The right panels show $\nu_e + n \rightarrow e^- + p$ while the right
panels show the reaction $\nu_e + n \rightarrow \nu_e' + n$. 
The dotted black
lines shows the position of the energy transfer peak when final state blocking
is not included, $U_4-U_2+M_4^*-M_2^*$. The solid lines show our full expressions for 
the cross section, the dot dashed lines show the full expressions neglecting tensor 
corrections, and the dashed lines show the \cite{Reddy:98} prescription for the cross 
section.}
\label{fig:dsigdq0}
\end{figure*}

\subsection{Differential Cross Sections}
\label{sec:differential}

The main results of this work are differential charged current neutrino
cross-sections. In general, the electron scattering angle and energy transfer are integrated
over to give the mean free paths which are relevant to transport calculations.
Nevertheless, to give the reader a feeling for the structure of these results, 
we show here some representative differential cross sections. 

First, we consider the structure of the full double differential absorption.
In figure \ref{fig:dsigdq0dmu}, we show the this quantity
at low density and at high density. It is strongly peaked at the $q_0$
expected for small momentum transfer. The variation with scattering angle is
given by the leading order $1/q = 1/\sqrt{E_1 + E_3 - 2 E_1 E_3 \mu_{13}}$
dependence of the baryon response combined with the lepton kinematics. Based on
the elastic limit of the cross section we expect a weak angular dependence in
the lepton kinematics of the form $1 + (C_V^2 - C_A^2)/(4C_V^2 + 12
C_A^2)\mu_{13} \approx 1 - 0.11 \mu_{13}$. The combination of these two factors
gives the strongly forward peaked differential cross section with a more slowly
varying backward scattering tail.
This leading order behavior only holds for neutrinos with energies small
compared to the baryon mass. At high density these limiting expressions for the
scattering angle dependence also break down and the structure of the
differential cross section is altered. For the conditions shown in the right
panel of \ref{fig:dsigdq0dmu}, it is slightly backward peaked and has strength
at a much larger range of energy transfers. 

Second, we consider the angle integrated differential cross section
$d\sigma/dq_0$, which encodes the energy transfer between the leptons
and baryons. For charged current reactions, we expect this quantity to be peaked
at $U_4 - U_2 + M^*_4 - M^*_2$ when electron final state blocking is ignored, since
this is the most favorable energy for small momentum transfer \cite{Roberts:12b,
Martinez-Pinedo:12}. In figure \ref{fig:dsigdq0}, this cross section is shown. 
The left panel shows the charged current differential cross-section at low density. 
Here, the electrons are not strongly degenerate and the peak of the differential
cross section is at the zero momentum transfer value for all but the smallest
neutrino energies shown. In contrast, at the high density shown in the right
panel, the differential cross section peaks away from the zero momentum transfer
value because the term $1-f_3(E_3)$ depends exponentially on $q_0$. Therefore,
there is significantly stronger dependence on the value of the differential
cross section away from the zero momentum transfer peak, which is often more
poorly captured in approximations to the charged current rates.

We can also start to see how the various corrections included in the rates
derived above alter the differential cross section. There, the tensor
corrections impact the normalization of the cross-section significantly,
increasing to a correction of $\sim 25 \%$ for the highest energy neutrinos, but
they do not impact the energy transfer. At high density, where $\Delta U$ is
larger and electron final state blocking can be significant, the variation
between different approximations to the charged current rates can be large. It
is also clear that the expressions of \cite{Reddy:98} differ significantly from
those derived here. We consider these differences in detail in the next section.

\begin{figure}[t] 
\includegraphics[width=0.49\textwidth]{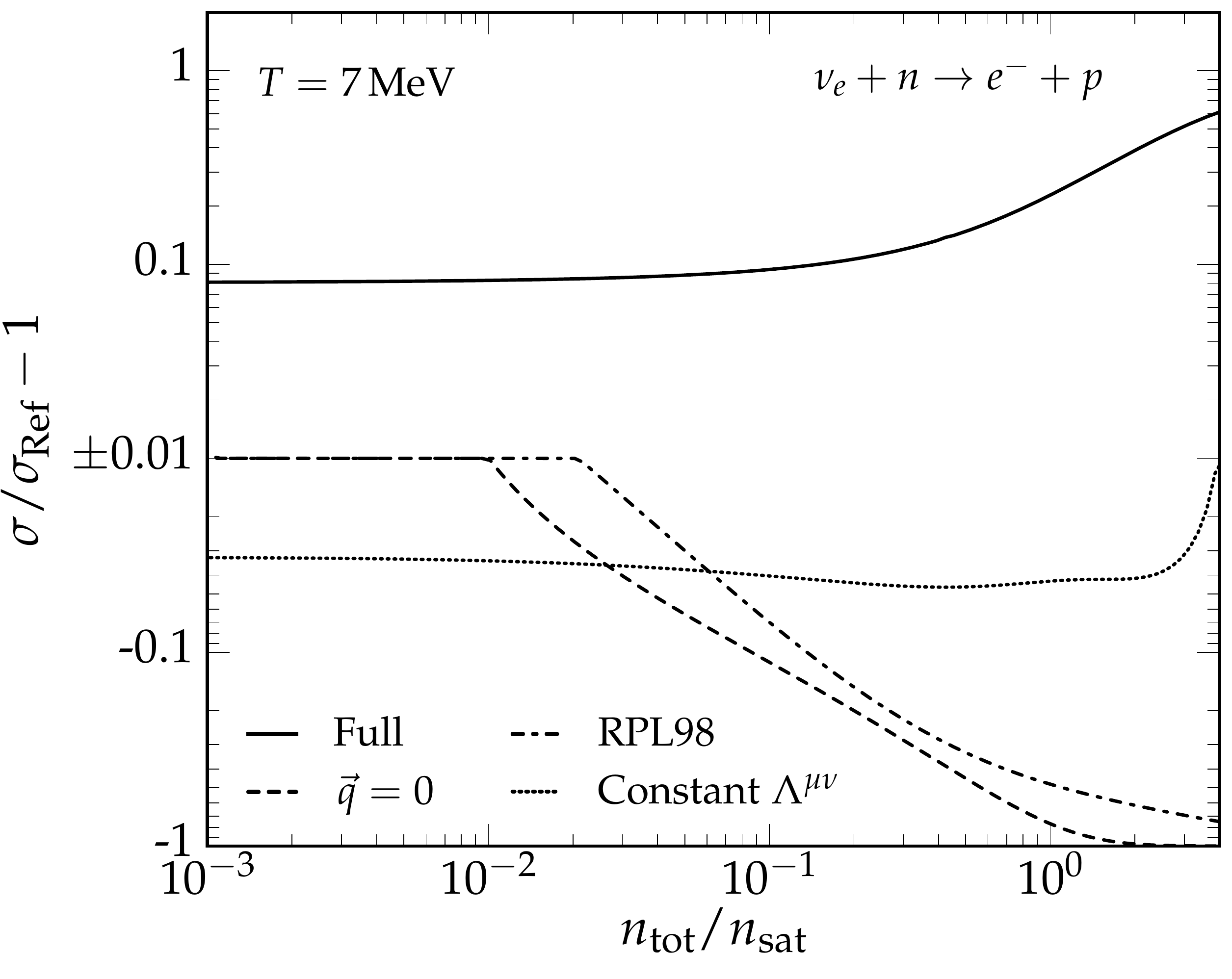} 
\includegraphics[width=0.49\textwidth]{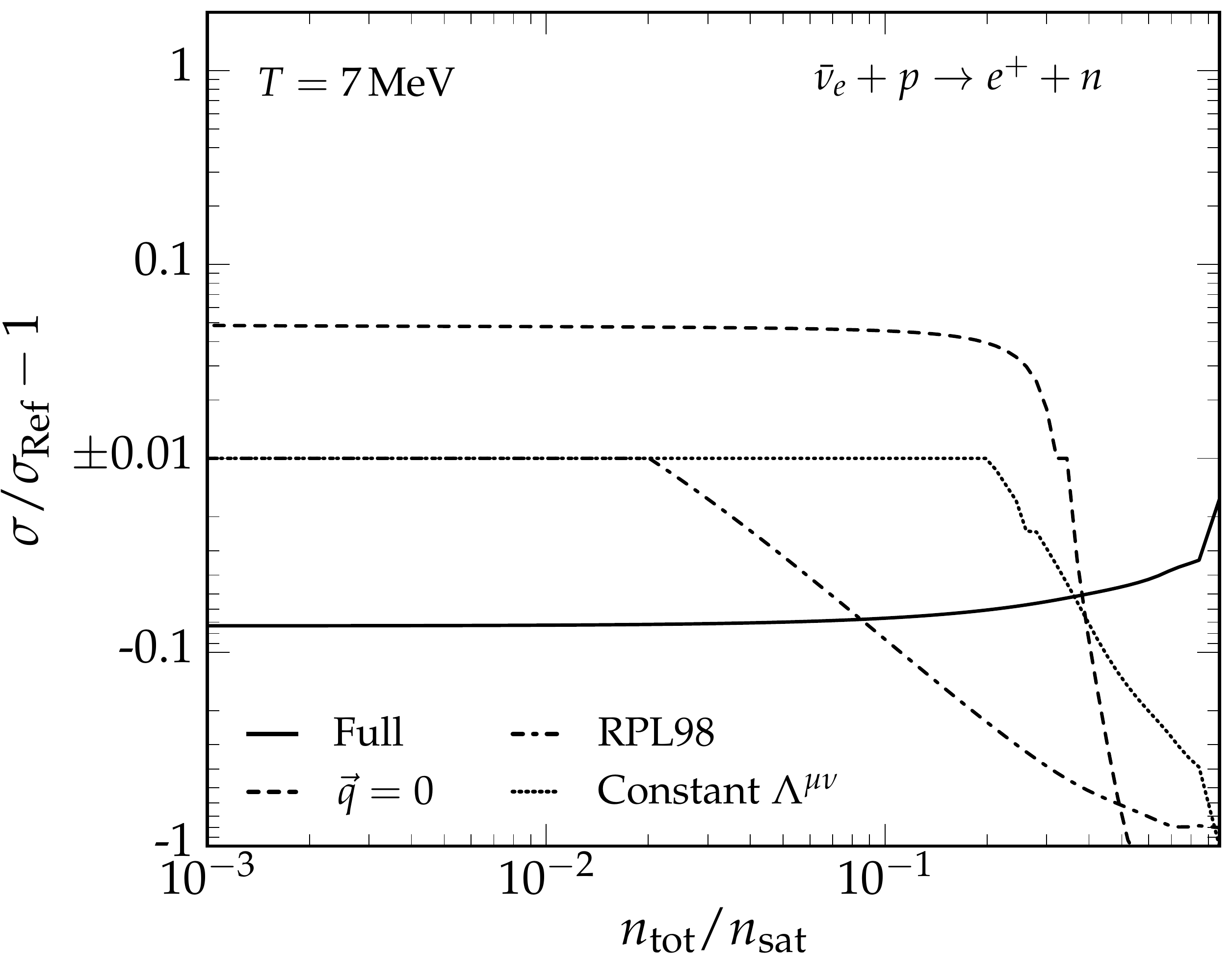} 
\caption{ Comparison of various approximations to the charged current
cross section per nucleon as a function of density in beta-equilibrated matter.
The reference cross section, $\sigma_{\rm Ref}$ is taken to be the full cross
section without weak magnetism corrections (i.e. $F_2 = 0.0$). The neutrino
energy is assumed to be $\pi T$. The left panel
shows the reaction $\nu_e + n \rightarrow e^- + p$ while the right panel shows 
$\bar \nu_e + p \rightarrow e^+ + n$. The electron antineutrino cross sections
are only shown up to threshold. Full denotes the fully relativistic cross
section including weak magnetism corrections, $\vec{q}=0$ is the cross section
given by Eq.~\ref{eq:csec_qeq0}, RPL98 denotes the cross section given
by the expressions in \cite{Reddy:98}, and Constant $\Lambda^{\mu\nu}$ refers to
the cross section given by Eq.~\ref{eq:csec_constant_M}.}
\label{fig:MFP_variation}
\end{figure}

\subsection{Comparison with earlier work}
\label{sec:comparison}

We now turn to compare our results for the charged current cross sections in neutron-rich nuclear matter including weak magnetism to a number of
other approximations for different ambient conditions. 
The comparisons are shown in figure \ref{fig:MFP_variation}. We take the full
expression with $F_2 = 0.0$ (i.e. the full cross section without weak magnetism
corrections) as a baseline for comparison.  The simplest approximation to the
charged current rates including mean field effects is given by equation
\ref{eq:csec_qeq0}. At low density, this approximation agrees with the full
cross section without weak magnetism at the percent level for $\nu_e + n
\rightarrow e^- + p$. The deviation is somewhat larger for electron antineutrino
capture.  Once the density becomes high enough for final state blocking to be
important, this approximation starts to strongly deviate from the full
expression and drastically under predicts the cross section at saturation
density and above for both electron neutrinos and antineutrinos. These
conclusions are likely to be impacted when the effective masses
have a more complex density dependence.  

The second approximation we consider comes from assuming a constant hadronic
portion of the matrix element (see Eq.~\ref{eq:csec_constant_M}). This
results in expressions that are very similar to the non-relativistic results
given by \citep{Reddy:98, Burrows:98}. At low density, this approximation agrees
with our baseline result to a few percent. This is not surprising given that
this approximation should be accurate to order $q/M$. At higher densities, this
approximation breaks down for the antineutrino capture rate.   

The inclusion of the full weak magnetism correction induces corrections of order
10\% for the 22 MeV neutrinos considered in the plot, with the correction going
in opposite directions for neutrino and antineutrino capture. Near saturation
density, the weak magnetism corrections begin to become significantly larger
than would be predicted by the expansion given in \cite{Horowitz:02}. 

\begin{figure}
\includegraphics[width=0.49\textwidth]{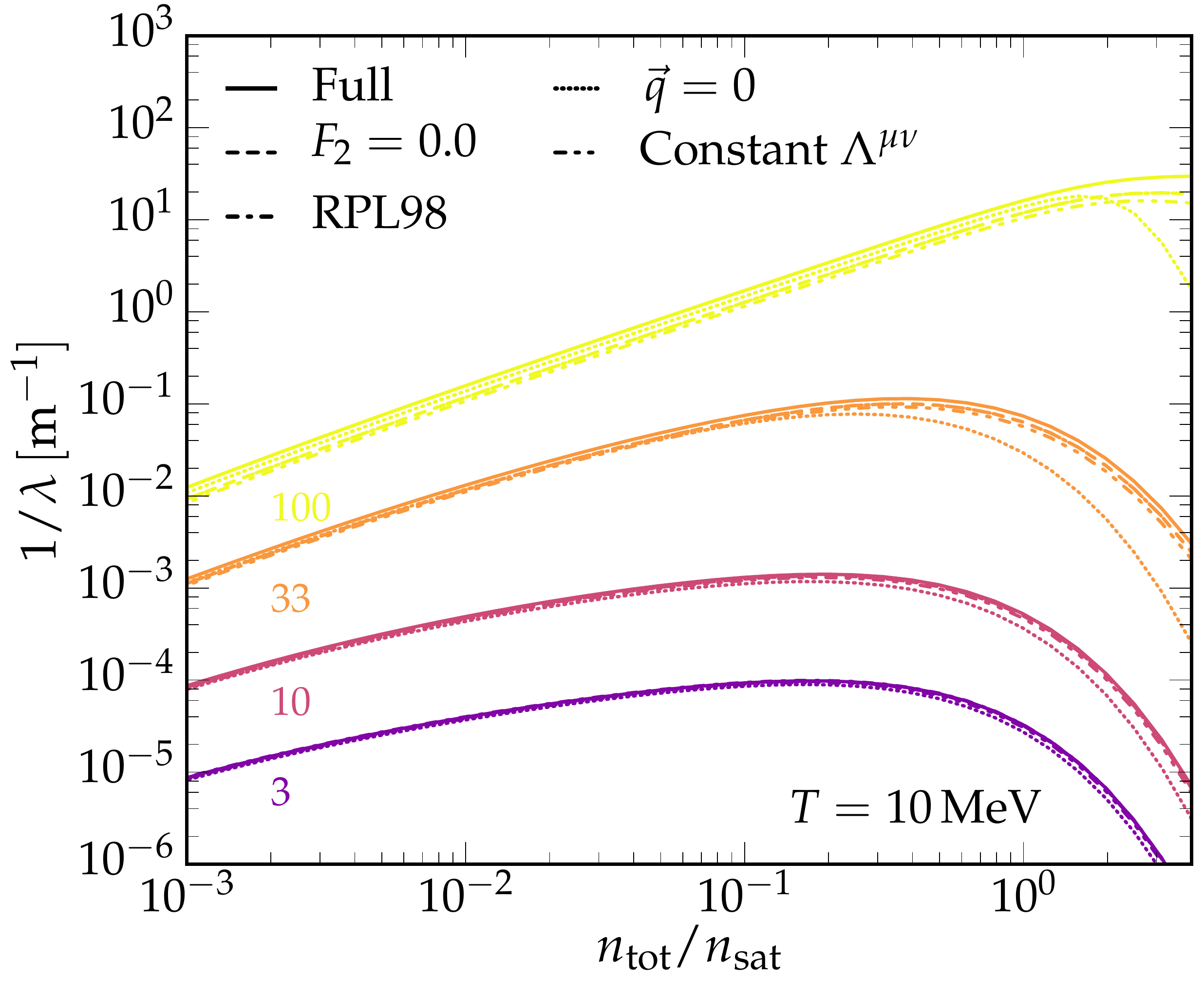} 
\includegraphics[width=0.49\textwidth]{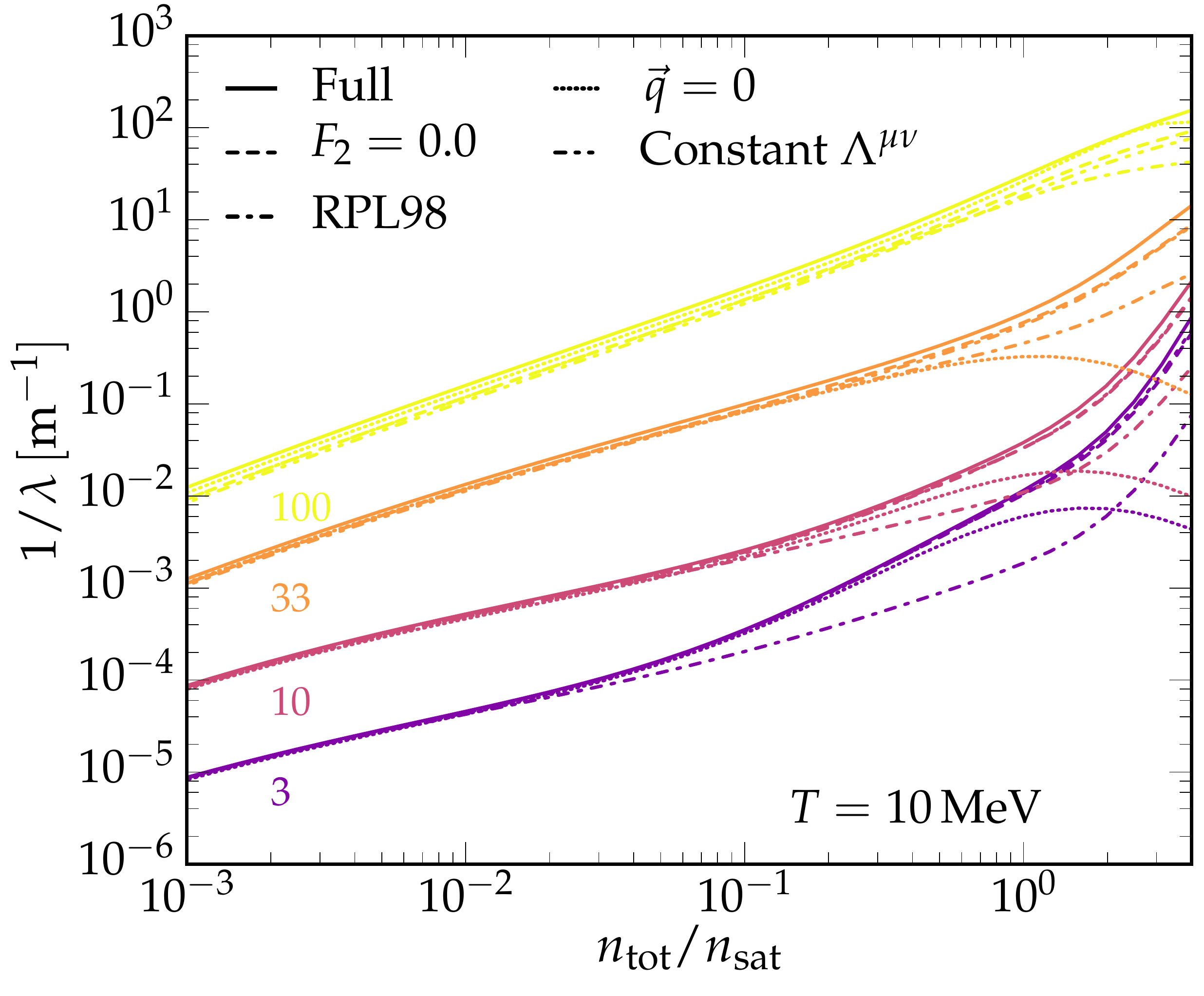} 
\caption{The $\nu_e + n \rightarrow e^- + p$ mean free path as a function of
total density for neutrino energies $\{3, 10, 33, 100\} \, \textrm{MeV}$.
The full cross section (solid lines), the full cross section neglecting weak
magnetism (dashed lines), the \cite{Reddy:98} expression for the cross section
(dot-dashed lines), the non-relativistic elastic limit (dotted lines), and the
constant $\Lambda_{\mu \nu}$ matrix element limit (dot-dot-dashed lines) of the
cross section are all shown. The left panel shows the results for a free gas of
protons and neutrons, while the right panel shows the results for a gas with a
mean field potential as described in the text. At low density, all of the
approximations agree reasonably well (although there are significant deviations
for higher energy neutrinos), but at higher density and for finite $\Delta U$
they diverge significantly.}
\label{fig:CC_MFP}
\end{figure}

A similar description of the relativistic charged current
differential neutrino cross sections was given in \cite{Reddy:98}. We find the following
significant differences from that work:  
\begin{enumerate}
\item The difference between neutron and proton masses and self-energies was not
properly accounted for in the calculation of the polarization tensor.
Specifically, the integrals needed to calculate the polarization function
defined in eqns.~\ref{eq:IsQ}, \ref{eq:IsL}, \ref{eq:IsM}, and \ref{eq:IsT}
depend on $\beta$ and $\tilde{q}_0$. These dependencies were neglected in
\cite{Reddy:98}  where $\beta=1$ and $\tilde{q}_0=q_0$ was used \footnote{This  appears
to be a typographical error because the code used to generate these results in
\cite{Reddy:98} did account for corrections arising from $\beta\neq1$ and
$\tilde{q}_0 \neq q_0$.}. 
\item Current conservation, which requires $q^\mu \Pi^{V}_{\mu,\nu}=0$, was used
to related different components of the polarization tensor in \cite{Reddy:98}.
Differences between the neutron and proton masses and self-energies violates
this relation \cite{Leinson:01}, and contracting the vector piece of the
polarization with the mean field corrected energy momentum transfer gives 
\begin{equation}
\tilde q^\mu \pp_{\mu \nu}^V = \pp^V_Q \tilde q^\nu + \pp_{M+}^V \tilde n^\nu \neq 0\,,
\label{eq:current_conservation}
\end{equation}
and since $ \pp_{\mu \nu}^V \propto \Pi^{V}_{\mu \nu}$, $q^\mu \Pi^{V}_{\mu
\nu}\neq 0$ in general. These corrections become significant when $U_2-U_4$ is
large compared to the neutrino energy. 
%\item Tensor charged currents were neglected in \cite{Reddy:98}, and although
%they contribute at the level of a few percent they are important because they
%enhance the cross section for electron neutrinos relative to the
%anti-neutrinos. 
\item There is typographical error in Eq. 69 and 70 of Ref. \cite{Reddy:98}
where the lower limit of the energy integral was defined. The correct expression
defined here in Eq.~\ref{eq:eminus1} differs by a sign. 
\item As discussed in \ref{sec:lepton}, the lepton tensor in \cite{Reddy:98} was
calculated by setting the charged lepton mass to zero because it was assumed
that their energies would be relatively large.  The associated correction for
reactions involving electron neutrinos is typically negligible, since $E_e \gg
m_e$ but can be important for charged current reactions involving muon
neutrinos.   
\end{enumerate}
These differences significantly alter the charged current cross sections in
regions where the $\Delta U$ is significant, as can be seen in 
figure~\ref{fig:MFP_variation}.

In figure~\ref{fig:CC_MFP}, the absorption mean free path in beta-equilibrated
matter is shown for a range of neutrino energies in all of the approximations
described above. In the left panel, we show the mean free path in a
non-interacting medium. At low density, all of the different expressions for the
capture rate agree reasonably well, with deviations increasing with increasing
neutrino energy. At high density, the cross section per nucleon is suppressed
due to final state electron blocking. This causes the $\vec{q}=0$ cross section for
higher energy neutrinos to be in error by factors of a few because of its delta
function distribution of allowed energy transfer, since the full
nucleon response is broadly peaked in $q_0$ for these neutrinos (see above
in section \ref{sec:differential}). All of the other approximations for the
cross section agree at the few tens of percent level across all densities for a
non-interacting gas.  The inclusion of weak magnetism corrections
only has a small impact, with the size of the correction increasing with
neutrino energy \cite{Horowitz:03}. 

When interactions are included, the differences between the various
approximations become more substantial, as was described above.
Additionally, the inclusion of mean field potentials has a strong impact of the
mean free paths relative to the free gas case \citep{Roberts:12b,
Martinez-Pinedo:12}. This is shown in the right panel of
figure~\ref{fig:CC_MFP}. At high density and low neutrino energy, the largest
deviations between the results in \cite{Reddy:98} and our results are seen. The
density at which significant deviations begin increases with increasing neutrino
energy. For smaller neutrino energies, $\Delta U$ becomes the dominant energy
scale at lower density than for higher energy neutrinos. Similarly to the
non-interacting case, the $\vec{q}=0$ results strongly under predict the inverse
mean free path at high density when final state blocking becomes larger. As
we saw above, constant $\Lambda_{\mu \nu}$ approximation agrees with the full
expression without weak magnetism quite well, although the deviations between
the two get larger with increasing neutrino energy. We expect the differences
would also get larger in models where the nucleon effective mass is
significantly less than the nucleon rest mass and other energy scales entering
the response function become closer to the in-medium nucleon mass. Finally, it
can be seen the corrections from the inclusion of weak magnetism are similar to
those in the non-interacting case.

\subsection{Nucleon Effective Masses} 
\label{sec:eff_mass}

\begin{figure}[t] 
\includegraphics[width=0.49\textwidth]{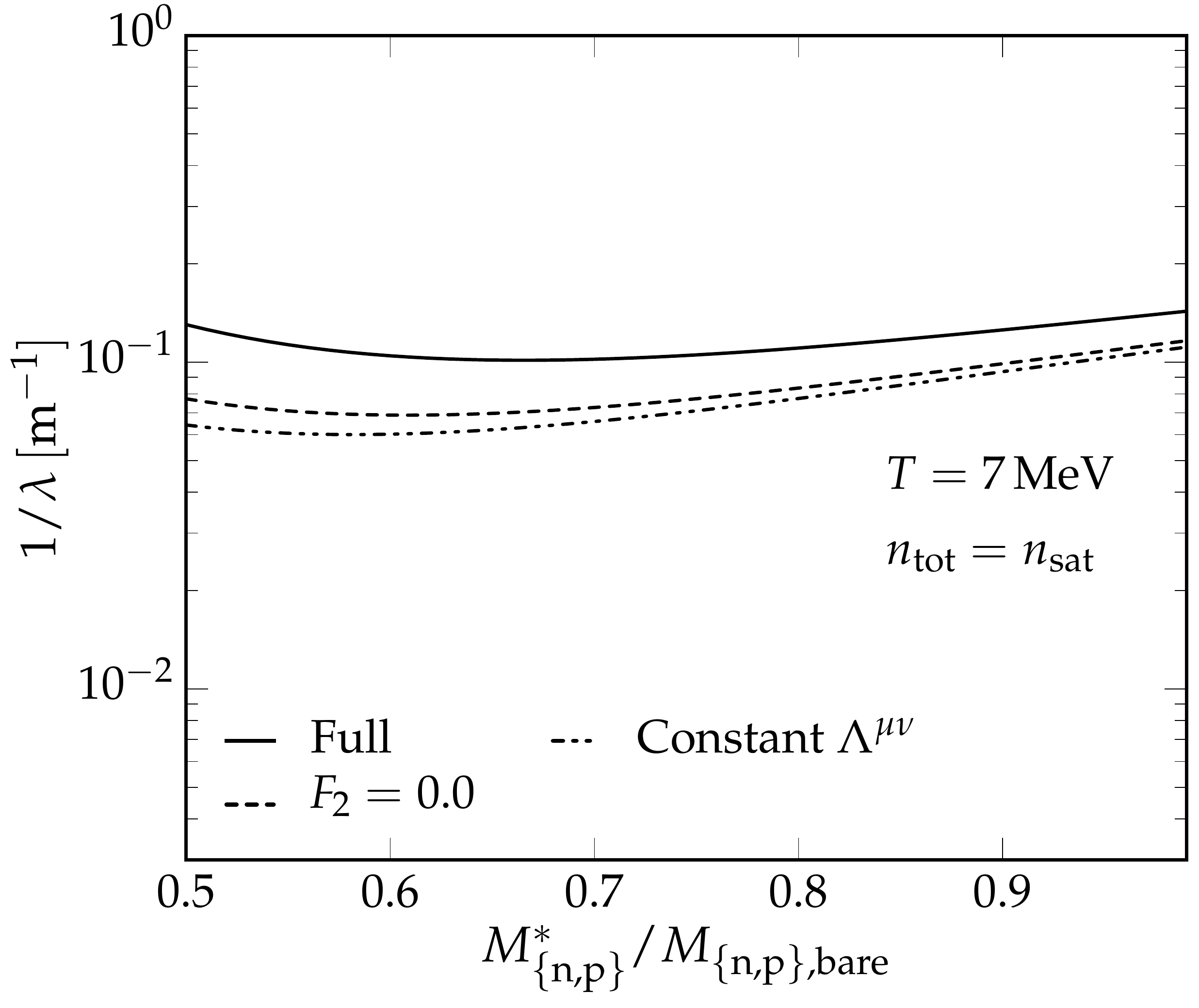} 
\caption{  Variation of the charged current neutrino mean free path with the
nucleon effective masses at saturation density. The 
nucleon effective mass is assumed to be isospin independent and matter is
assumed to be in beta-equilibrium. The neutrino energy is taken to be $E_\nu =
\pi T$. Full denotes the fully relativistic cross section including weak
magnetism corrections, RPL98 denotes the cross section given by the expressions
in \cite{Reddy:98}, and Constant $\Lambda^{\mu\nu}$ refers to the cross section
given by Eq.~\ref{eq:csec_constant_M}. } 
\label{fig:eff_mass} 
\end{figure}

In the illustrative calculations above, we kept the nucleon masses fixed at
their vacuum values for simplicity. In the nuclear medium, nucleon effective
masses can be significantly different from their bare values.  In
non-degenerate matter, we expect the dependence of the mean free path on the
isospin independent effective mass to be weak \cite{Reddy:99}. On the other
hand, in degenerate matter the
leading order $M^*_2 M^*_4$ dependence of the differential cross-section 
remains after integration over $q_0$ and $\mu_{13}$ \cite{Reddy:99}. Therefore,
we expect that reducing the nucleon effective mass will reduce the neutrino mean
free path but at the a rate slower than $M^2$.  In figure \ref{fig:eff_mass}, we show the impact of varying the
nucleon effective masses at saturation density in beta-equilibrium assuming the
effective masses are isospin independent. The expected reduction in the mean
free path is seen in all approximations of the neutrino mean free path down to
$M^*/M \approx 0.6$. Because the chemical potentials in beta-equilibrium depend
on the effective mass, $\hat \mu$ decreases with increasing effective mass in Figure
\ref{fig:eff_mass}. Therefore, the detailed balance factor gets larger for smaller
effective masses and eventually becomes more important than the leading order mass
dependence in the response, causing the turnover in the behavior of the mean
free path with effective mass seen at the lowest effective masses considered
here. 

It is also likely that there is significant isospin dependence of the effective
mass \cite{Li:15}. The formalism developed here can fully account for this. To
leading order, isospin splitting of the effective mass provides a similar effect
to isospin dependent potential energies, since the charged current response
will be peaked around $q_0 = U_4 - U_2 + M^*_2 - M^*_4 = 0$. Including this
isospin dependence in the effective masses in a manner consistent with
constraints on the nuclear symmetry energy requires also choosing different
values for the nucleon potential energies. A detailed investigation of the
impact of isospin dependent effective masses on neutrino opacities is beyond the
scope of this work.

\subsection{The Neutral Current Limit}
 Since the neutral current
interactions have the same mass and potential for the incoming and outgoing
baryons, we set $\beta=1$, $\lambda=\Delta=0$, $\sigma_- = 1$, $\sigma_+ = 
1 - 4 m_2^2/q_\alpha^2$, and $\tilde q_0 = q_0$. 
With these replacements, only two pieces of the vector polarization are non-zero  
\begin{eqnarray}
\pp_L^{V,nc} &=& \pp_Q + \pp_L \nonumber \\
&=& -\frac{q_\mu^2}{4 \pi q^3} \int_{e_m}^\infty dE_2 (f_2 - f_4) 
\left[(2 E_2 + q_0)^2 - q^2 \right] \\
\pp_T^{V,nc} &=& \pp_Q + \pp_T
\nonumber \\
&=& -\frac{q_\mu^2}{4 \pi q^3} \int_{e_m}^\infty dE_2 (f_2 - f_4)  
\left[q^2/2 + 2 M_2^2 \frac{q^2}{q_\alpha^2} + (2E_2 + q_0)^2/2 \right] 
\end{eqnarray}
which has the same form as the results of \cite{Reddy:98} (although there is a
factor of two difference, which is purely definitional).  
For the neutral current reactions, vector current conservation holds for the polarization 
tensor, such that $q_\mu \pp_V^{\mu \nu}= 0$, which can be seen from
substituting the neutral current expressions in Eq.~\ref{eq:current_conservation}. 
Therefore,
the standard decomposition of the vector part of the neutral current polarization 
tensor is
\begin{equation}
\pp_{\mu\nu}^{V,nc} = \pp_L^{V,nc} P^L_{\mu \nu} + \pp_T^{V,nc} P^T_{\mu \nu},
\end{equation}
which gives $\pp^{V,nc}_L = -\frac{q_\mu^2}{q^2} \pp^{V,nc}_{00}$ and $\pp^{V,nc}_T =
-\pp^{nc}_{22}$, which has been used in previous works \citep{Reddy:98, Horowitz:12}.

We can also see that 
\begin{equation}
\pp^{A,nc}_{\mu \nu} = \pp^{V,nc}_L P^L_{\mu \nu} + \pp^{V,nc}_T P^T_{\mu \nu} 
+ \eta_{\mu \nu} \pp_A
\end{equation}
with 
\begin{equation}
\pp_A  = -\frac{4 M_2^2}{q_\alpha^2} \pp_Q
= -\frac{M_2^2}{\pi q} \int_{e_m}^\infty dE_2 (f_2 - f_4),
\end{equation}
which also agrees with \cite{Reddy:98}. Our expression for the mixed vector-axial polarization is
also in agreement. 

The neutral current tensor polarizations were investigated in 
\cite{Horowitz:03}. Our result is 
\begin{eqnarray}
\pp^T_L &=& \pp_Q - \frac{q_\alpha^2}{4 m_2^2} \pp_L =-\frac{q_\alpha^2}{4 m_2^2}\left[\pp^V_L 
+ \pp_A \left(1+\frac{q_\alpha^2}{4 m_2^2}\right) \right]
\\
\pp^T_T &=& \pp_Q - \frac{q_\alpha^2}{4 m_2^2} \pp_T =-\frac{q_\alpha^2}{4 m_2^2}\left[\pp^V_T 
+ \pp_A \left(1+\frac{q_\alpha^2}{4 m_2^2}\right) \right]
\end{eqnarray}
This agrees with \cite{Horowitz:03} up to a sign in front of $\pp_A$ in
$\pp^T_T$.

\section{Conclusions}
We have derived complete expressions for charged current neutrino interactions
in the mean field approximation, including weak magnetism, arbitrary degeneracy, and
relativistic kinematics.  We approach the problem both using a Fermi's Golden
Rule and the linear response given by many-body perturbation theory to clarify the derivation and provide a path to
including correlations. Both approaches of course yield the same
answer. We then investigated the neutrino mean free paths predicted by these
results and compared them to other results found in the literature.  
These expressions can also be used to calculate neutral current cross
sections and inelastic electron scattering. We provide an open source
library for calculating these opacities at \codewebsite.

Our results extend and correct results previously presented in the literature.
We find that vector current non-conservation in charged current reactions
\cite{Leinson:01} and the correct inclusion of neutron and proton potential
energies, along with weak magnetism corrections to all orders, introduces a
number of new terms in the charged current opacities.
At low densities, the corrections to the old rates are modest and provide
changes of a few to ten percent relative to previous results in the literature
for neutrino energies of order ten MeV.
This is unlikely to make qualitative changes in the results of core collapse
supernova and compact object merger simulations, but
in situations where computational models are sensitive to small changes in
microphysics these corrections can at least remove one source of uncertainty.
At high density, especially when the neutron proton self-energy difference is
large, they can alter the neutrino mean free paths in the medium by factors of a
few or greater. This is likely to impact the rate of lepton transport inside
proto-neutron stars.

Additionally, our derivation of the charged current polarization tensor provides
a starting point for future calculations of RPA corrections to the charged
current rates.  Previous work has found that these corrections can change rates
by up to factors of a few at and above nuclear saturation density
\citep{Burrows:99, Reddy:99}. At sub-nuclear density, more work is warranted to
assess the role of particle-hole screening and collective modes. Since the
nucleon-nucleon interaction is nearly resonant at low momentum, we are exploring
the RPA correlations with large effective interactions that can reproduce
nucleon-nucleon phase shifts. This will be reported in a future publication.
Additionally, because we find a different tensor structure than was found in
previous work \citep{Burrows:99, Reddy:99}, the detailed form of the RPA
equations will be altered. It remains to be seen if this significantly impacts
the resulting opacities. In the future, we also plan to include RPA corrections
in our publicly released opacity library.

\begin{acknowledgments}
We gratefully acknowledge Ermal Rrapaj for pointing out the chemical potential
difference appearing in the Bosonic frequencies of the imaginary time
polarization and Vincenzo Cirigliano for many useful discussions.  We also thank
the participants of the INT workshop INT-16-61W for useful discussions that
highlighted the importance of charged current reactions in supernovae.     

This research was supported by the NSF under TCAN grant number AST-1333607.
S. R. was supported by the US Dept. of Energy Grant No. DE-FG02-00ER41132.  
This work was enabled in part by the NSF under Grant No. PHY-1430152 (JINA
Center for the Evolution of the Elements). 
\end{acknowledgments}

\appendix 
\section{Polarization Tensor Decomposition} 
\label{app:Decomposition}

Here, we partially follow \cite{Le-Bellac:00} on pages 113, 118, and 214. First,
we choose to decompose the momentum transfer as $\tilde q_\mu = (\tilde q_0, 0,
0, q)$ and build a second vector orthogonal to $\tilde q_\mu$ in the time-z plane,
$n_\mu = (q, 0, 0, \tilde q_0)$. We then define the transverse projector 
\begin{eqnarray}
P_{\mu \nu}^{T+} &=& \eta_{\mu \nu} 
- \frac{\tilde q_\mu \tilde q_\nu}{\tilde q_\mu^2} 
- \frac{n_\mu n_\nu}{n^2} 
\nonumber\\
&=& \eta_{\mu \nu} + \frac{n_\mu n_\nu - \tilde q_\mu \tilde q_\nu}{\tilde q_\mu^2},
\end{eqnarray}
which picks out the portion of a vector orthogonal to $\tilde q_\mu$ and $n_\mu$. We 
can also define the projectors along $\tilde q_\mu$ and $n_\mu$ as
\begin{eqnarray}
P^L_{\mu \nu} &=&  \frac{n_\mu n_\nu}{n_\mu^2} = -\frac{n_\mu n_\nu}{\tilde q_\mu^2}  \\
P^Q_{\mu \nu} &=&  \frac{\tilde q_\mu \tilde q_\nu}{\tilde q_\mu^2}. 
\end{eqnarray}
It is also useful to define the quantities 
\begin{eqnarray}
P^{M\pm}_{\mu \nu} &=&  \frac{\tilde q_\mu n_\nu \pm \tilde q_\nu n_\mu}{\tilde q_\mu^2} \\
P^{T-}_{\mu \nu} &=&  \epsilon_{\mu \nu \lambda \delta} 
\frac{\tilde q^\lambda n^\delta}{\tilde q_\alpha^2}
\end{eqnarray} 
The only non-zero, complete contractions of these tensors with one another are 
\begin{eqnarray}
P^L_{\mu \nu}P_L^{\mu \nu} &=& 1 \\ 
P^Q_{\mu \nu}P_Q^{\mu \nu} &=& 1 \\ 
P^{M\pm}_{\mu \nu}P_{M\pm}^{\mu \nu} &=& -2 \\
P^{T\pm}_{\mu \nu}P_{T\pm}^{\mu \nu} &=& 2.
\end{eqnarray}

Any rank two tensor can be decomposed  as 
\begin{equation}
A_{\mu\nu} = A_Q P^Q_{\mu \nu} + A_L P^L_{\mu \nu} 
+ A_{M+} P^{M+}_{\mu \nu} + A_{M-} P^{M-}_{\mu \nu} + \tilde A_{\mu \nu}.
\end{equation}
If there are no other preferred directions that enter into $\tilde A_{\mu \nu}$
(which is the case with the polarization tensors as long as the distribution
function is isotropic), we have 
\begin{equation}
\tilde A_{\mu \nu} = A_{T+} P^{T+}_{\mu \nu} + A_{T-} P^{T-}_{\mu \nu},
\end{equation} 
since $P^T$ is the only available transverse, symmetric tensor and the second 
piece is the only anti-symmetric tensor that can be made from the Levi-Civita 
tensor. They can be recovered using 
\begin{eqnarray}
A_Q &=& P_Q^{\mu \nu} A_{\mu \nu}\\
A_L &=& P_L^{\mu \nu} A_{\mu \nu}\\ 
A_{M+} &=&-\frac{1}{2} P_{M+}^{\mu \nu} A_{\mu \nu}\\
A_{M-} &=&-\frac{1}{2} P_{M-}^{\mu \nu} A_{\mu \nu}\\
A_{T+} &=&\frac{1}{2} P^{\mu \nu}_{T+} A_{\mu \nu} = -A_{22} \\
A_{T-} &=&\frac{1}{2} P^{\mu \nu}_{T-} A_{\mu \nu}.
\end{eqnarray}
The last relation on the second to last line holds when $n_\mu$ and $q_\mu$ are 
in the 0-3 plane. 

\section{Spin Sums in RMF Theory} 
\label{app:MeanField}
Here we derive the spin sums of Dirac spinors essential in defining the nucleon propagators in mean field theory given earlier in Eq.~\ref{eq:G_MF}. 
For Walecka type field theories, we generally have a Lagrangian of the the form 
\begin{eqnarray}
\script{L} &=& \sum_B \bar \Psi_B (i \gamma^\mu \partial_\mu - M_B 
+ g_{\sigma B} \sigma - g_{\omega B} \gamma^\mu \omega_\mu)\Psi_B +
\script{L}_{\sigma \omega}
\end{eqnarray}
where we have only included a single vector field $\omega_\mu$ for simplicity
and the meson field self-interactions are subsumed into $\script{L}_{\sigma
\omega}$. The addition of isospin dependent fields to this Lagrangian is a
trivial extension of the results presented below. 
In the mean field approximation, the scalar and vector fields
behave classically and for homogeneous matter the only non-zero component of the
vector field is the time component. 
For this Lagrangian, the equations of motion 
\begin{eqnarray}
(i \gamma^\mu \partial_\mu - M_B + g_{\sigma B} \sigma 
- g_{\omega B} \gamma^\mu \omega_\mu)\Psi_B = 0\,, \\
\bar \Psi_B(-i \gamma^\mu \ \overleftarrow{\partial}_\mu - M_B 
+ g_{\sigma B} \sigma - g_{\omega B} \gamma^\mu \omega_\mu) = 0\,,
\end{eqnarray}
are the analog of the Dirac equation in the presence of classical background fields. 

First, we can find the dispersion relations nucleons by noting that fields that satisfy the  
Dirac equation also satisfy the Klein-Gordon equation. In standard notation, we
then have 
\begin{eqnarray}
(i \fsl{\partial} - g_{\omega B} \fsl{\omega} - M^*_B)(-i \fsl{\partial} +
g_{\omega B} \fsl{\omega}
 - M^*_B) \Psi_B
= \left[(i \partial_\mu - g_{\omega B} \omega_\mu)(i \partial^\mu - g_{\omega B}
\omega^\mu) - {M_B^*}^2 \right] \Psi_B=0
\end{eqnarray}
which admits plane wave solutions $\Psi_B \propto
\exp(\mp i p \cdot x)$, and the dispersion relation is given by the equation for $p^\mu$ 
\begin{equation}
(p_\mu \mp g_{\omega B} \omega_\mu)(p^\mu \mp g_{\omega B} \omega^\mu) - {M_B^*}^2 = 0.
\end{equation}
Similar to the free field case, the upper sign will correspond to particle while
the lower sign will correspond to anti-particles and from now on we will only
consider the particles. The equation for $\bar \Psi_B$ gives a similar result and if we define the four-vector 
\begin{equation}
\tilde p^\mu = p_\mu - g_{\omega B} \omega^\mu = (E^*, -\vec{p}),
\end{equation}
the mean field theory will look almost exactly like the free field theory, just
with the replacement $p^\mu \rightarrow \tilde p^\mu$ everywhere except for in
the exponent.  Here $E^* = \sqrt{p^2 + {M_B^*}^2}$ is the kinetic energy of the
particle.

With this spatial dependence, we can expand the baryon fields in terms of Fourier modes and
promote the expansion coefficients to creation and annihilation operators $a_s$
and $b_s$ and the standard four-component spinners $u_s$ and $v_s$ as 
\begin{eqnarray}
\Psi_B(x) &=& \sum_{s=+,-} \int \frac{d^3p}{(2 \pi)^3 2 E_p^*} 
\left(a_s(\vec{p}) u_s(\vec{p}) e^{-ip\cdot x} + b^\dagger_s(\vec{p}) v_s(\vec{p}) 
e^{ip\cdot x} \right ) \\ 
\bar \Psi_B(x) &=&  \sum_{s=+,-} \int \frac{d^3p}{(2 \pi)^3 2 E_p^*} 
\left(a_s^\dagger(\vec{p}) \bar u_s(\vec{p}) e^{ip\cdot x} + b_s(\vec{p}) 
\bar v_s(\vec{p}) e^{-ip\cdot x} \right ).
\end{eqnarray}
Enforcing the standard anti-commutation relations on the fields gives 
\begin{equation}
\left\{a^\dagger_s(\vec{p}), a_{s'}(\vec{p}') \right\} = 
(2 \pi)^3 \delta^{(3)}(\vec{p} - \vec{p}') 2 E^*_p \delta_{ss'},
\end{equation}
where the $E^*_p$ normalization is consistent with the denominator of the
Lorentz invariant phase space factor in the field expansions. 

It is easy to show that the Hamiltonian density is 
\begin{equation}
\script{H} = \bar \Psi_B \left(i \gamma^j \partial_j + m + 
g \fsl{\omega} \right) \Psi_B.
\end{equation}  
Integrating over space and using the field expansions gives the Hamiltonian 
\begin{equation} 
H = \sum_s 
\int \frac{d^3p}{(2 \pi)^3 2 E^*_p} 
\left[E^*_p a^\dagger_{p,s} a_{p',s'} \right] + g \omega_0 Q +
\textrm{anti-particles},  
\end{equation}
where 
\begin{equation}
Q = \int \frac{d^3p}{(2 \pi)^3 2 E^*_p} \left[a^\dagger_{p,s} a_{p',s'} \right] 
- \textrm{anti-particles},
\end{equation} 
is the baryon number. This result is useful when considering the grand canonical
Hamiltonian, which gets also gets a contribution $-\mu Q$.  

We also need to know how the presence of mean fields impacts the properties of
the four-component spinors.  Using the field equation for the baryon fields with
the above expansions results in
\begin{eqnarray}
( \fsl{p} - M_*  - g_{\omega B} \fsl{\omega})u_s(\vec{p}) = 0 \\
\bar u_s(\vec{p})  ( \fsl{p} - M_*  - g_{\omega B} \fsl{\omega})= 0 \\
( -\fsl{p} - M_* - g_{\omega B} \fsl{\omega})v_s(\vec{p}) = 0 \\
\bar v_s(\vec{p}) ( -\fsl{p} - M_* - g_{\omega B} \fsl{\omega})= 0 
\end{eqnarray}
The equations for $u_s(\vec{p})$ can be expressed in terms of $\tilde p^\mu$, in
which case they take the free field form.  This allows us to write down the
spin sums 
\begin{eqnarray}
\sum_{s=\pm} \bar u_s(\vec{p}) u_s(\vec{p}) &=& \fsl{\tilde p} + M_* \\
\sum_{s=\pm} \bar v_s(\vec{p}) v_s(\vec{p}) &=& \fsl{\tilde p} - M_*,
\end{eqnarray}
which appear in the mean field propagator.  Additionally, this implies that the
Gordon identities and other properties of the spinors are unaltered from the
free field case aside from the aforementioned replacement.  

We emphasize that the above analysis dealt only explicitly with particle states. 
The anti-particles get a potential energy with the opposite sign from the mean
field, so that the anti-particle kinetic four-momentum is given by $\tilde p_\mu
= p_\mu + q \omega_\mu$. 

\section{Matsubara Sums} 
\label{app:Matsubara}

Here, we calculate the types of Matsubara sums necessary to evaluate the
imaginary time polarization tensor.  It is easiest to consider the Matsubara sums
directly. Using standard methods \citep[e.g.][]{Le-Bellac:00}, it is easy 
to show that 
\begin{eqnarray}
 T \sum_n g(i \omega_n + \alpha) = \sum \textrm{Res} \, g(p_0) f(p_0 - \alpha),  
\end{eqnarray}
where $g$ is an arbitrary function of a complex variable that converges more
rapidly than $1/|p_0|$ in all directions (otherwise it is necessary to be
careful about which convergence factor is chosen). The pair bubble will require
evaluating Matsubara sums of the form 
\begin{eqnarray}
S^{l,j}(i\omega_m + \Delta \mu) &=& T \sum_n 
(i \omega_n + \nu_1)^l(-i (\omega_m - \omega_n) + \nu_2)^j \nonumber \\
&& \quad \times \Delta(i \omega_n + \nu_1, E_1) 
\Delta(i (\omega_m + \omega_n) + \nu_2, E_2),
\end{eqnarray}
where $\Delta(i\omega, E) = 1/(\omega^2 + E_p^2)$ and 
$E_i = \sqrt{p_i^2 + m_i^2}$.  We choose the energy of a particle of
species 2 (rather than an anti-particle), since we will primarily be interested
in the particle-particle contribution to the polarization. This results in a $g$
of the form
\begin{widetext}
\begin{eqnarray} 
g(p_0) =  \frac{p_0^l (p_0 - \omega_0)^j}
{[E_1^2 - p_0^2][E_2^2 - (\omega_0 - p_0)^2]} 
= \frac{p_0^l (p_0 - \omega_0)^j}
{(E_1 + p_0)(E_1 - p_0)(E_2 + \omega_0 - p_0)(E_2 - \omega_0 + p_0)},
\end{eqnarray}
where $\omega_0 = -i \omega_m + \Delta \nu$ and $\omega_m$ is a Bosonic frequency.
The relations 
\begin{equation}
\frac{1}{(a+b)(a-b)} = \frac{1}{2a}\left[\frac{1}{a-b} + \frac{1}{a+b} \right]
= \frac{1}{2b}\left[\frac{1}{a-b} - \frac{1}{a+b} \right]
\label{eq:pole}
\end{equation}
help define the properties of the poles found in $g$. 
The $g$ function has four simple poles. Using the notation of equation
\ref{eq:pole}, we find.
\begin{enumerate} 

\item $p_0 = E_1$, $a = E_2$, and $b = E_1 - \omega_0$ 
\begin{eqnarray*}
\textrm{Res} \, g(E_1) f &=& -\frac{E_1^l (E_1 - \omega_0)^j}{4 E_1 E_2}
\left[\frac{f(E_1 - \nu_1)}{\omega_0 - E_1 + E_2} 
- \frac{f(E_1 - \nu_1)}{\omega_0 - E_1 - E_2} \right] 
\end{eqnarray*}

\item $p_0 = -E_1$, $a = E_2$, and $b = E_1 + \omega_0$  
\begin{eqnarray*}
\textrm{Res} \, g(-E_1) f &=& -\frac{(-E_1)^l (-\omega_0 - E_1)^j}{4 E_1 E_2}
  \left[\frac{1 - f(E_1 + \nu_1)}{\omega_0 + E_1 - E_2} 
- \frac{1 - f(E_1 + \nu_1)}{\omega_0 + E_1 + E_2} \right] 
\end{eqnarray*}

\item $p_0 = \omega_0+E_2$, $a = E_1$, and $b = E_2 + \omega_0$
\begin{eqnarray*}
\textrm{Res} \, g(\omega_0 + E_2) f &=& -\frac{(\omega_0 + E_2)^l E_2^j}{4 E_1 E_2}
 \left[-\frac{f(E_2 - \nu_2)}{\omega_0 - E_1 + E_2} 
+ \frac{f(E_2 - \nu_2)}{\omega_0 + E_1 + E_2} \right] 
\end{eqnarray*}

\item $p_0 = \omega_0-E_2$, $a=E_1$, and $b=E_2 - \omega_0$
\begin{eqnarray*}
\textrm{Res} \, g(\omega_0 - E_2) f &=& 
\frac{( - E_2 + \omega_0)^l (-E_2)^j}{4 E_1 E_2}
\left[\frac{1 - f(E_2 + \nu_2)}{\omega_0 + E_1 - E_2} 
- \frac{1 - f(E_2 + \nu_2)}{\omega_0 - E_1 - E_2} \right] 
\end{eqnarray*}
\end{enumerate}
Combining terms that have the same denominators and carefully accounting for signs, 
we then have (for $l,j = \{0,1\}$)
\begin{eqnarray}
S^{l,j}(i \omega_m + \Delta \mu) &=&
-\frac{E_1^l E_2^j}{4 E_1 E_2} \biggl[ 
\frac{f_1(E_1) - f_2(E_2)}{-i \omega_m + \Delta \nu - E_1 + E_2} 
+(-1)^j \frac{1 - f_1(E_1) - \bar f_2(E_2)}{-i \omega_m + \Delta \nu - E_1 - E_2} \nonumber\\
&& - (-1)^{l+j}\frac{\bar f_1(E_1) - \bar f_2(E_2)}{-i \omega_m + \Delta \nu + E_1 - E_2} 
- (-1)^l \frac{1 - \bar f_1(E_1) - f_2(E_2)}{-i \omega_m + \Delta \nu + E_1 + E_2} \biggr],
\end{eqnarray} 
where $f_i(E) = [\exp(\beta E - \beta \nu_i)+1]^{-1}$.
\end{widetext}

\end{document}